\documentclass[aps,preprint,onecolumn,floatfix,nofootinbib,superscriptaddress]{revtex4-1}

\usepackage{amsmath}
\usepackage{amssymb}
\usepackage{amsfonts}
\usepackage{float}
\usepackage{color}
\usepackage{mathtools}                                  
\usepackage{bbold}                                      
\usepackage{listings}                                   
\usepackage{bm}                                         
\usepackage{subfig}                                     

\DeclareMathOperator{\Tr}{Tr}
\DeclareMathOperator{\erf}{erf}

\newcommand{\rom}[1]{\uppercase\expandafter{\romannumeral #1\relax}}    

\makeatletter
\def\l@subsection#1#2{}
\def\l@subsubsection#1#2{}
\makeatother

\begin{document}

\pagenumbering{arabic}

\title{Analytically projected rotationally symmetric explicitly correlated Gaussian functions with one-axis shifted centers}

\author{Andrea Muolo}
\affiliation{ETH Z\"urich, Laboratory of Physical Chemistry, Vladimir-Prelog-Weg 2, 8093 Z\"urich, Switzerland}

\author{Markus Reiher}
\email{Corresponding author: markus.reiher@phys.chem.ethz.ch}
\affiliation{ETH Z\"urich, Laboratory of Physical Chemistry, Vladimir-Prelog-Weg 2, 8093 Z\"urich, Switzerland}

\date{June 8, 2020}

\begin{abstract}

A new explicitly correlated functional form for expanding the wave function of an $N$-particle system 
with arbitrary angular momentum and parity is presented. 
We develop the projection-based approach, numerically exploited in our previous work 
[J. Chem. Phys. {\textbf{149}}, 184105 (2018)], 
to explicitly correlated Gausssians with one-axis shifted centers and derive 
the matrix elements for the Hamiltonian and the angular momentum operators 
by analytically solving the integral projection operator. 
Variational few-body calculations without assuming the Born-Oppenheimer approximation 
are presented for several rotationally excited states of three- and four-particle systems.
We show how the new formalism can be used as a unified framework for high-accuracy calculations 
of properties of small atoms and molecules.

\end{abstract}

\maketitle

\section{Introduction}\label{SEC:intro}

Highly accurate bound states of the Schr\"odinger equation for small atoms and molecules can be constructed
by expanding the wave function in terms of basis functions depending explicitly on inter-particle distances
\cite{ECG-history1960,ECG-history1960_1,svm-history1977,randTempe-1,FECG-integrals-advances1978,Szalewicz_1979,Szalewicz_II_1983,randTempe-2,randTempe-3,Szalewicz_1990,ECG-history1993,Drake1997,Korobov2000,Adamowicz2003a,Matyus2012,Pachucki2012,Adamowicz2013,Pachucki2015}.
Non-separable functions with respect to the particle coordinates are tailored to describe particle-particle correlations, 
especially to accurately reproduce the exact wave function for infinitesimally short distances and in the long range limit. 
Furthermore, they allow for a unified treatment of different kinds of particles, {\it{e.g.}} of electrons and nuclei. 
Within this framework, two- and three-electron atoms can be very accurately calculated employing Hylleraas-type functions
\cite{Hylleraas1928,Hylleraas1929a,Hylleraas1929b,Hylleraas1930a,Hylleraas1930b,Drake1997}
that explicitly include powers of the inter-electronic distances $r_{ij}=|\bm{r}_i-\bm{r}_j|$.
However, the difficulties of the analytical calculation of their matrix elements prevent application 
of this approach to larger systems 
\cite{Perkins1973,Hyl3ele_1987,Langer2011}.
Generality with respect to the particle number and accessible analytical Hamiltonian matrix elements 
are achievable through powers of the quadratic form of the inter-particle distances 
that define explicitly correlated Gaussian-type (ECG) functions \cite{ECG-history1960,ECG-history1960_1}.
Plain explicitly correlated Gaussian (pECG) functions for $N_p$ interacting particles
\begin{align}
\phi_I^{\text{pECG}} = \exp\left[-\sum_{i<j=1}^{N_p}{A_I}_{ij}\bm{r}_i\cdot\bm{r}_j\right] ~,
\end{align}
are the simplest functions of this type and have been successfully employed 
to describe a number of diverse physical systems, from small atoms and molecules to light nuclei,
hadrons, quantum dots, and Efimov systems \cite{Matyus2012,Adamowicz2013,Adamowicz2013_rev}.
pECG functions are also manifestly spherically symmetric, i.e. invariant under rotation, 
as they are eigenfunctions of the total angular momentum squared operator with eigenvalue zero. 
Additional and important higher angular momentum contributions
originate from the cross terms of the exponential part, i.e. $\exp(-{A_I}_{ij}\bm{r}_i\cdot\bm{r}_j)$ which,
when expanded into a power series, contain terms of the form 
\begin{align}
(\bm{r}_i\cdot\bm{r}_j)^n = \sum_{2k+l=n} \frac{4\pi(2k+l)!}{2^kk!(2k+2l+1)!!} 
~ {|r_i|}^{2k} {|r_j|}^{2k} 
\sum_{m=-l}^{l} \mathcal{Y}_{lm}(\bm{r}_i) \mathcal{Y}_{lm}(\bm{r}_j) ~,
\end{align}
which are associated with different solid spherical harmonics $\mathcal{Y}_{lm}$ 
for the coordinates $\bm{r}_i$ and $\bm{r}_j$.

Although these advantages
made ECG-type functions very popular in high accuracy calculations 
\cite{svm-history1977,FECG-integrals-advances1978,ECG-history1993,Varga1995,Korobov2000,Adamowicz2003a},
the spherical symmetry limits the applicability of plain ECGs to ground rotational states only. 
Different approaches \cite{Varga1996,Adamowicz2013_rev} 
have been developed to extend ECGs to nonspherical problems, 
i.e. for calculating states with non-zero total spatial angular momentum quantum numbers $N$.  

In general, the ECGs are being multiplied with a nonspherical function $\theta_{NM_N}(\bm{r})$ of the collective position 
vectors $\bm{r}$
that for one particle in a central potential would just reduce to a solid spherical harmonic $\mathcal{Y}(\bm{r}_1)$.
The generalization to the $N_p$-particle case is a vector-coupled product of the solid spherical harmonics of the relative coordinates,
\begin{align}
\theta_{NM_N}(\bm{r}) = \sum_{\kappa=\{m_1,m_2,\ldots,m_{N_p}\}} \mathcal{C}_{\kappa} \prod_{i=1}^{N_p} \mathcal{Y}_{l_im_i}(\bm{r}_i) ~,
\label{eq:partWaveDecomp}
\end{align}
where $\mathcal{C}_{\kappa}$ is a product of Clebsch--Gordan coefficients,
\begin{align}
\mathcal{C}_{\kappa} =& \langle l_1m_1l_2m_2|L_{12}m_1+m_2\rangle \langle L_{12}m_1+m_2l_3m_3|L_{123}m_1+m_2+m_3\rangle \nonumber \\
& \cdots \langle L_{12\ldots N_p-1}m_1+m_2+\ldots+m_{N_p-1}l_{N_p}m_{N_p}|NM_N\rangle ~,
\end{align}
that couples the orbital angular momenta sequentially to the specified total quantum numbers ($N,M_N$).
Since the angular momentum of the relative motion is not a conserved quantity, it is important for an accurate description to include
several sets of orbital angular momenta ($l_1,l_2,\ldots,l_{N_p};L_{12},L_{123},\ldots$) weighted by $\mathcal{C}_{\kappa}$.
Eq.~(\ref{eq:partWaveDecomp}) is a partial-wave expansion whose direct implementation is cumbersome
since the matrix elements for this choice of $\theta_{NM_N}(\bm{r})$ will become very complicated. 
Moreover the algebraic complexity of the integral matrix elements is not invariant with respect to the number of particles, 
and hence, analytical expressions must be derived for each different system.

One viable alternative to the full partial wave decomposition is to consider only limited coupling schemes  
``specializing'' the basis functions for a given $N$ while the relative matrix elements are explicitly derived.
For example, Refs.~\cite{Komasa2001,Adamowicz2008_zCECG,Adamowicz2009_N1,Adamowicz2013_N1,Adamowicz2015_zECG} 
focused on ECG functions specifically tailored for $N=1$ states considering
the sets of orbital angular momenta ($l_1=0,\ldots,l_i=1,\ldots,l_{N_p}=0$).
Ref.~\cite{Adamowicz2010_N2,Adamowicz2011_N2,Adamowicz2011_N2_Ryd1,Adamowicz2011_N2_Ryd2})
tackled $N=2$ states analogously with lowest-order angular momentum couplings. 

Alternatively, representations of $\theta_{NM_N}(\bm{r})$ including the orientation of a global vector $\bm{v}$
formed as a linear combination of all particle coordinates $\{\bm{r}_i\}$, have been successfully employed 
in high-accuracy calculations of properties of small atoms and molecules
\cite{suzukivarga1998,Matyus2012}.
This approach is based on an equivalence condition between the global vector representation of $\theta_{NM_N}(\bm{r})$ 
and the partial-wave expansion for a given orientation of the global vector.
Under the assumption of a smooth energy landscape in parameter space, 
the global vector orientation can be recovered variationally through the minimization of the energy
with respect to its real-valued parameters. 
Although this approach is appealing because it yields analytical matrix elements for quantum mechanical operators
that are form invariant with respect to the angular momentum quantum numbers $N$ and $M_N$, 
and the number of particles $N_p$,
the variational optimization of the global vector parameters is difficult and not every $\theta_{NM_N}(\bm{r})$ can be represented.
These alternative formulations are strictly derived from the partial wave expansion as a result of having truncated or  
variationally approximated Eq.~(\ref{eq:partWaveDecomp}).

In this work, we extend our numerical projection scheme onto irreducible representations of the rotational-inversion O(3) group 
presented in our previous work \cite{Muolo2018b},
focusing on a special case where the integral projector can now be solved analytically.
In Ref.~\cite{Muolo2018b},
we considered explicitly correlated Gaussian functions with centers shifted 
by a vector in the three dimensional Euclidean space, $\bm{s}\in\mathbb{R}^3$.
Numerically exact eigenfunctions of the squared total spatial angular momentum operator $\hat{\bm{N}}^2$ and 
the parity operator $\hat{p}$ were then constructed with explicit projection onto the corresponding eigenspace.
We relied on numerical quadrature schemes for the calculation of integral matrix elements which 
introduced noticeable computational cost in the variational iterative steps.
In practice, numerical projection precludes large basis sets from being optimized variationally
and limits the applicability of the developed formalism. 
Here, we consider solving exactly the projection operator for a subset of floating ECG functions
having shifted centers along only one axis. 
We devise analytical integral matrix elements for projected functions for the overlap, kinetic, Coulomb, and angular momentum operators. 
We illustrate the validity of this novel functional form by studying the first three rotational states of the dihydrogen molecular 
ion, H$_2^+=\{$p$^+,$p$^+,$e$^-\}$ treated explicitly as a three-particle system
and the dyhydrogen molecule, H$_2=\{$p$^+,$p$^+,$e$^-,$e$^-\}$ treated explicitly as a four-particle system.

\section{Theory} \label{SEC:theory}

We consider a non-relativistic Coulombic Hamiltonian for $N_p$ particles 
\begin{equation}
\label{nonrel-H}
\hat{H}_{\text{lab}} = -\bm{\nabla_{r}}^T M \bm{\nabla_r} + 
\sum_{i=1}^{N_p} \sum_{j>i}^{N_p} \frac{q_iq_j}{\left|\bm{r}_i-\bm{r}_j\right|} ~,
\end{equation}
with the position vector $\bm{r}_i$ of the $i$th particle 
in the laboratory fixed Cartesian coordinates (LFCC), its mass $m_i$ and its charge $q_i$.
$\bm{\nabla_r}$ is the gradient with respect to $\bm{r}_i$ and $M$ is a $N_p\times N_p$ matrix with elements $M_{ij}=\delta_{ij}/2m_i$.

As we are interested in bound states, the motion of the center of mass (CM) can be discarded. 
This is usually realized by a linear transformation of the coordinates 
\begin{equation}
\label{prima-transf}
U_x\bm{r}=\left(\bm{x}_1,\bm{x}_2,\ldots,\bm{x}_{N_p-1},\bm{x}_{\text{CM}}\right)^T
\end{equation}
in which the $\bm{x}_{\text{CM}}=\sum_{i=1}^{N_p}m_i\boldsymbol{r}_i/(\sum_{i=1}^{N_p}m_i)$
are the center-of-mass Cartesian coordinates and $\bm{x}\equiv(\bm{x}_1,\ldots,\bm{x}_{N_p-1})$
denotes the translationally invariant Cartesian coordinates (TICC) corresponding to the internal coordinates of the system
generated through the relative tranformation matrix $U_x$.
A transformation of the Hamiltonian in Eq.~(\ref{nonrel-H}) separates the kinetic energy term 
for the center of mass from the internal Hamiltonian \cite{suzukivarga,Adamowicz2013_rev}:
\begin{align}
\hat{H}_{\text{int}} = - \nabla_{\bm{x}}^\text{T}\,\mu\,\nabla_{\bm{x}} +
\sum_{i=1}^{N_p-1} \sum_{j>i}^{N_p-1} \frac{q_iq_j}{|(\bm{f}_{ij}\otimes\mathbb{1}_3)\bm{x}|} ~,
\label{eq:TI-Hamiltonian}
\end{align}
where 
\begin{align}
\mu =& U_x^{-T}MU_x ~,
\end{align}
and
\begin{align}
(\bm{f}_{ij})_k =& (U^{-1}_x)_{ik} - (U^{-1}_x)_{jk} ~.
\end{align}
This separation of the center-of-mass coordinate requires transforming both the Hamiltonian and the state function 
and has been exploited in practice \cite{Adamowicz2003a,Matyus2012}.

By contrast, here we solely transform the basis functions in a given TICC set without transforming quantum mechanical operators
following the method described in our previous work \cite{Benjamin2013,Muolo2018a}.
In this approach, the matrix-element calculations are carried out naturally in the LFCC set and the center-of-mass contamination 
is rigorously subtracted from the expectation values.
While handling state functions in a TICC set is very appealing because of the restriction of the parameter space to only 
$N_p-1$ internal coordinates, we avoid the difficulties arising from matrix elements for transformed operators and instead
retain the algebraic simpler and intuitive LFCC set for the integral evaluation. 
We employ the heavy-particle centered, the center-of-mass centered, and Jacobian Cartesian coordinate sets, 
allowing the basis functions to cycle through these TICC representations in order to describe efficiently 
different ''groupings`` of particles (e.g., pairs and triples of particles).

\section{Basis functions} \label{SEC:basisfunction}

Given the total spin quantum number and its projection on the $z$-axis $S$ and $M_s$, respectively, 
the wave function representing is expanded as a linear combination of (anti-)symmetrized 
floating explicitly correlated Gaussian (FECG) functions
\begin{equation}
\label{eq:wavefun}
\Psi(\mathbf{r})=\sum_{I=1}^{N_b}c_{I} \, \bm{\chi}_{I}^{S,M_{S}} \, \hat{Y}\phi_{I}^{{\text{FECG}}}
(\bm{r};A_I^{(r)},\bm{s}_I^{(r)}) ~,
\end{equation}
where $c_I$ are the expansion coefficients, $\bm{\chi}_{I}^{S,M_{S}}$ are spin functions, and $\hat{Y}$ 
is the Young operator that accounts for the appropriate permutation symmetry of sets of identical particles 
as described by Kinghorn \cite{Kinghorn1996}.
FECGs have the following general form
\begin{align}
\phi_I^{\text{FECG}} (\bm{r};A_I^{(r)},\bm{s}_I^{(r)})
&= \exp\left[-(\bm{r}-\bm{s}_I^{(r)})^T (A_I^{(r)}\otimes\mathbb{1}_3) (\bm{r}-\bm{s}_I^{(r)}) \right] ~.
\label{eq:fecg}
\end{align}
Here, $A_I^{(r)}$ 
is an $N_p\times N_p$ symmetric matrix of the $\frac{1}{2}N_p(N_p+1)$ variational parameter, 
with the subscript $I$ indicating that the matrix is unique for each basis function
and the superscript indicating that the variational parameters refer to the LFCC set. 
It is $\bm{r}(A_I^{(r)}\otimes\mathbb{1}_3)\bm{r}>0 \,\,\forall\,\, \bm{r}\in\mathbb{R}^{3N_p}$,
that is $A_I^{(r)}$ must be positive definite,
to ensure square integrability of the $\phi_I^{[{\text{FECG}}]}$ basis function.
A necessary and sufficient condition for a symmetric real matrix to be positive definite is that all eigenvalues must be positive.
Here $\bm{r}-\bm{s}_I^{(r)}$ stands for a set of vectors 
$\{\bm{r}_1-\bm{s}_{I\,1}^{(r)},\ldots,\bm{r}_{N_p}-\bm{s}_{I\,N_p}^{(r)}\}$
that correspond to shifted particle coordinates with the $3N_p$-dimensional vector 
$\bm{s}_I^{(r)}$ composed of parameters to be optimized in a variational procedure.

Note that the floating spherical Gaussian orbitals (FSGO) approach introduced by Frost in 1967
\cite{Frost1967} is based on one-particle functions (orbitals) and
is therefore a limiting case of our approach for diagonal (and not dense)
Gaussian parameter matrices $A_I$.
In fact, this special case reduces our FECG basis functions to a product of exponential functions,
each of which being spherically symmetric about its origin.
By contrast, FECG basis functions with dense $A_I$ Gaussian parameter matrices,
include partial waves contributions from many higher angular momentum states (see the Introduction).

In the following sections we explicitly work out the integral matrix elements in the simple LFCC frame.

\section{Projection technique} \label{SEC:projection}

The FECGs in Eq.~(\ref{eq:fecg}) define Gaussian functions with shifted centers to allow for suitable deformations of the 
ansatz for the all-particle wave function
that are predominantly needed for polyatomic systems \cite{Adamowicz2013_rev,Muolo2018a}.
A general FECG function is, however, neither an eigenfunction of the squared
total angular momentum operator $\bm{N}^2$, nor an eigenfunction of the 
space inversion operator $\hat{p}$. 
As the rotation-inversion symmetry must be restored variationally in the limit of a complete basis set, 
these basis functions gives rise to poor energy convergence.

To alleviate this problem, we recently proposed an integral projection operator, $\hat{P}_{M_N}^{[N,p]}$ \cite{Muolo2018b}, 
to ensure the correct spatial rotation-inversion symmetry corresponding to $N$ and $M_N$, 
the total spatial angular momentum quantum numbers, and the parity quantum number $p$:
\begin{equation}
\label{eq:projOp}
\hat{P}_{M_N}^{[N,p]} = \hat{P}_{M_NM_N}^{[N]} \, \hat{P}_{C_I}^{[p]} ~,
\end{equation}
with
\begin{align}
\hat{P}_{M_1M_2}^{[N]} =& \int \frac{d\Omega}{4\pi^3} 
\,\, D^{[N]}_{M_1M_2}\left(\Omega\right)^* \hat{R}\left(\Omega\right) ~,
\end{align}
and
\begin{align}
\hat{P}_{C_I}^{[p]} =& \hat{\mathcal{E}} + p\cdot \hat{\mathcal{I}} ~,
\end{align}
where $\hat{\mathcal{E}}$ is the identity operator, $\hat{\mathcal{I}}$ is the spatial inversion operator,
and $D_{M_1M_2}^{[N]}$ is the element of the $N$-th Wigner $D$-matrix 
\begin{align}
D^{[N]}_{M_1M_2} = \exp(-iM_1\alpha) \, d^{[N]}_{M_1M_2}(\beta) \exp(-iM_2\gamma) ~,
\end{align}
with the Wigner (small) $d$-matrix being
\begin{align}
d^{[N]}_{M_1M_2}(\beta) =& \big[(N+M_1)!(N-M_1)!(N+M_2)!(N-M_2)!\big]^{\frac{1}{2}}
\nonumber \\[0.1cm]
& \times \sum_s \Bigg[ \frac{ (-1)^{M_1-M_2+s} \left(\cos\frac{\beta}{2}\right)^{2N+M_2-M_1-2s} 
\left(\sin\frac{\beta}{2}\right)^{M_1-M_2+2s} } { (N+M_2-s)!s!(M_1-M_2+s)!(N-M_1-s)! } \Bigg] ~.
\end{align}
$\hat{R}(\Omega)$ is the quantum mechanical rotation operator
over the Euler angles $\Omega\equiv\{\alpha,\beta,\gamma\}$ \cite{Rose:AngularMomentum},
\begin{align}
\hat{R}(\alpha,\beta,\gamma) = \exp(-i\alpha N_z) \exp(-i\beta N_y) \exp(-i\gamma N_z) ~.
\label{eq:RotOp1}
\end{align}
The effect of the projector operator in Eq.~(\ref{eq:projOp}) on a state $|N\,M_N\rangle$ is
\begin{align}
\hat{P}_{M_1M_2}^{[N_1]} |N_2M_2\rangle = |N_1M_1\rangle \,\delta_{N_1N_2} \,\delta_{M_1M_2} ~,
\end{align}
with $|NM_N\rangle$ being angular momentum eigenstates. 
Note that our original implementation \cite{Muolo2018b} of the projection scheme was purely numerical, which we overcome
in this work for the special case of projection on one spatial axis, for which an analytical expression can be derived.

The form of the rotation operators in Eq.~(\ref{eq:RotOp1}) is not a convenient operational definition because they require an explicit expression 
of the angular momentum components $N_i$ that is not entirely straightforward in our all-particle explicitly-correlated formulation.
Nonetheless, exactly the same symmetry operation will be realized if we rotate the physical system itself 
or if we rotate the coordinate axis in the opposite direction, 
\begin{align}
\hat{R}(\Omega)\phi^{{\text{FECG}}}_I\big(\bm{r};A_I^{(r)},\bm{s}_I^{(r)}\big) & = 
\phi^{{\text{FECG}}}_I\big(U(\Omega)^{-1}\,\bm{r};A_I^{(r)},\bm{s}_I^{(r)}\big) 
\nonumber \\
& = \exp\left[-\big(U(\Omega)^{-1}\,\bm{r}-\bm{s}_I^{(r)}\big)^T
\big(\bar{A}_I^{(r)}\otimes\mathbb{1}_3\big)
\big(U(\Omega)^{-1}\,\bm{r}-\bm{s}_I^{(r)}\big)\right] \nonumber \\
& = \exp\left[-\big(\bm{r}-U(\Omega)\bm{s}_I^{(r)}\big)^T
\big(\bar{A}_I^{(r)}\otimes \tilde{U}(\Omega)^{-T}\tilde{U}(\Omega)^{-1}\big)
\big(\bm{r}-U(\Omega)\bm{s}_I^{(r)}\big)\right] \nonumber \\
& = \phi_I^{{\text{FECG}}}\big(\bm{r};A_I^{(r)},U(\Omega)\bm{s}_I^{(r)}\big) ~,
\label{eq:FormInvariance}
\end{align}
where $U(\Omega)=\mathbb{1}_{N_p}\otimes\tilde{U}(\Omega)$
represents the coordinate transformation generalized to a system of $N_p$ particles with
\begin{align}
\tilde{U}(\Omega) = 
\left(\begin{array}{ccc} 
\cos\alpha\cos\beta\cos\gamma-\sin\alpha\sin\gamma & -\cos\gamma\sin\alpha-\cos\alpha\cos\beta\sin\gamma & -\cos\alpha\sin\beta \\ 
\cos\beta\cos\gamma\sin\alpha+\cos\alpha\sin\gamma & \cos\alpha\cos\gamma-\cos\beta\sin\alpha\sin\gamma  & -\sin\alpha\sin\beta \\ 
\cos\gamma\sin\beta & \sin\beta\sin\gamma & \cos\beta 
\end{array}\right) ~.
\label{eq:RotOp}
\end{align}

The properties of the rotation operator are summarized in four commutation relations:
\begin{align}
 &\left[\hat{R}(\Omega),\hat{H}\,\right]  \,   = \,\, 0 ~, \label{eq:projOpProp1} \\
 &\left[\hat{R}(\Omega),\hat{N}^2\right]       = \,\, 0 ~, \label{eq:projOpProp2} \\
 &\left[\hat{R}(\Omega),\hat{N}_z\right]  \ne    \,\, 0 ~, \label{eq:projOpProp3} \\
 &\left[\hat{R}(\Omega),\hat{p}\,\right]  \,   = \,\, 0 ~. \label{eq:projOpProp4}
\end{align}
Furthermore, the $\hat{P}_{M_NM_N}^{[N]}$ projection operator is idempotent and Hermitian:
\begin{align}
(\hat{P}_{M_NM_N}^{[N]})^2 =& \hat{P}_{M_NM_N}^{[N]} \label{eq:projOpProp5} \\
(\hat{P}_{M_NM_N}^{[N]})^{\dagger} =& \hat{P}_{M_NM_N}^{[N]} \label{eq:projOpProp6} ~.
\end{align}

Properties in Eqs.~(\ref{eq:projOpProp1})-(\ref{eq:projOpProp6}) are employed in the remainder of this work 
for the calculation of quantum mechanical expectation values.

\section{Matrix elements} \label{SEC:matrixelements}

In this section, we present analytically projected {FECG}s matrix elements for important operators
in the special case of unidimensional shift vectors, that is, employing $\bm{s}_I$ shift vectors of the form
\begin{equation}
\label{zFECG_0}
\bm{s}_I^{(r)} = \bm{u}_I^{(r)} \otimes \bm{e}_z ~,
\end{equation}
where $\bm{u}_I^{(r)}$ is a vector of length $N_p$ and $\bm{e}_z=\left(0,0,1\right)^T$.
From this choice of the $\bm{s}_I^{(r)}$ vectors we obtain the fundamental relation
\begin{equation}
\label{zFECG_0_1_fun2}
\bm{e}_z^T\tilde{U}(\Omega)\bm{e}_z = \cos\beta ~.
\end{equation}

Eq.~(\ref{zFECG_0_1_fun2}) is employed throughout this work to derive analytical matrix elements 
for the overlap, kinetic, Coulomb, and angular momentum operators.
For the matrix element of these operators we start from the analytical expressions 
derived for plain FECG by Cafiero and Adamowicz \cite{Adamowicz2002_FECG}.
Conversely, angular momentum matrix elements are derived from the analytical expressions 
for plain FECG presented in our previous work \cite{Muolo2018b}.
The unprojected and analytically projected $z$-shifted floating explicitly correlated 
Gaussian functions are abbreviated with zFECGs and apzFECGs,
respectively.

Given a quantum mechanical operator $\hat{O}$ commuting with the projector operator,
the matrix element $IJ$ for apzFECGs reads as follows:  
\begin{align}
\mathcal{O}_{IJ[N,M_N,p]}^{\text{apzFECG}} =
& \left\langle \phi^{\text{apzFECG}}_{I[N,M_N,p]} \big| \hat{O} \big| \phi^{\text{apzFECG}}_{J[N,M_N,p]} \right\rangle
= \left\langle \phi_I^{\text{zFECG}} \big| \hat{O} \big| \hat{P}^{[N,p]}_{M_N} \phi_J^{\text{zFECG}} \right\rangle ~,
\label{eq:MatEl}
\end{align}
where the Hermiticity and idempotency of the projection operator, Eqs.~(\ref{eq:projOpProp5}) and (\ref{eq:projOpProp6}), 
were exploited to simplify the integral expression.
In the following, analytical matrix elements for a variety of quantum mechanical operators are derived. 
For the sake of brevity, the projection onto the parity states $\hat{P}_{C_I}^{[p]}$ is omitted.

\subsection{Overlap integral}

The matrix elements of the identity operator for plain FECGs are given by \cite{Adamowicz2002_FECG}
\begin{align}
\left\langle \phi_I^{\text{FECG}} | \phi_J^{\text{FECG}} \right\rangle 
=\, \tilde{S}_{IJ} \, \exp\left[2 \bm{s}_I^{(r)^T}
A_I^{(r)} A_{IJ}^{{(r)}^{-1}} A_{J}^{(r)} \bm{s}_J^{(r)}\right] ~,
\label{eq:overlap_plainFECG}
\end{align}
where $A_{IJ}^{(r)}=A_I^{(r)}+A_J^{(r)}$ and
\begin{align}
\tilde{S}_{IJ} =& \left(\frac{\pi^{N_p}}
{\left|\bar{A}_I^{(r)}+\bar{A}_J^{(r)}\right|}\right)^{\frac{3}{2}} 
\exp\left[- \bm{s}_I^{(r)^T} A_I^{(r)} \bm{s}_I^{(r)}
-\bm{s}_J^{(r)^T} A_J^{(r)} \bm{s}_J^{(r)}\right] \nonumber \\
& \times \exp\left[+ \bm{s}_I^{(r)^T} A_I^{(r)} A_{IJ}^{{(r)}^{-1}}
A_{I}^{(r)} \bm{s}_I^{(r)}
+\bm{s}_J^{(r)^T} A_J^{(r)} A_{IJ}^{(r)^T}
A_{J}^{(r)} \bm{s}_J^{(r)}\right] ~.
\end{align}

In Eq.~(\ref{eq:overlap_plainFECG}) we have separated $\tilde{S}_{IJ}$, 
the term unaffected by the action of the rotation operator on the shift vector $\bm{s}_J^{(r)}$.
The remaining term must be investigated since it involves the angular integration over the Euler angles.
For apzFECGs the overlap matrix element reads
\begin{align}
S^{\text{apzFECG}}_{IJ[N,M_N,p]} 
=& \left\langle \phi_I^{\text{zFECG}} (\bm{r};A_I^{(r)},\bm{s}_I^{(r)}) \big| 
\hat{P}^{[N,p]}_{M_N} \phi_J^{\text{zFECG}} (\bm{r};A_J^{(r)},\bm{s}_J^{(r)}) \right\rangle ~,
\end{align}
and writing explicitly the projection operator leads to
\begin{align}
\label{eq:SapzFECG}
S_{IJ[N,M_N,p]}^{\text{apzFECG}}
=& \int \frac{d\Omega}{4\pi^3} \,\, D^{[N]}_{M_NM_N}(\Omega)^* 
\left\langle \phi_I (\bm{r};A_I^{(r)},\bm{s}_I^{(r)} \big| 
\phi_J (\bm{r};A_J^{(r)},U(\Omega)\bm{s}_J^{(r)} ) \right\rangle ~,
\end{align}
where we again drop the projector onto the parity state for the sake of brevity.
Because $\tilde{S}_{IJ}$ is invariant under the action of $\hat{P}^{[N,p]}_{M_N}$, Eq.~(\ref{eq:SapzFECG}) can be written as
\begin{align}
S_{IJ[N,M_N,p]}^{\text{apzFECG}} = \, \tilde{S}_{IJ} \, \Upsilon^{N}_{M_N} ~,
\end{align}
with
\begin{align}
\Upsilon^{N}_{M_N} = \int \frac{d\Omega}{4\pi^3} \,\, 
D^{[N]}_{M_NM_N}(\Omega)^* \, \exp\left[2\, \bm{s}_I^{(r)^T} A_I^{(r)} A_{IJ}^{{(r)}^{-1}}
A_{J}^{(r)} U(\Omega) \bm{s}_J^{(r)} \right] ~.
\label{zFECG_0_1}
\end{align}
Since $U(\Omega)=\mathbb{1}_{N_p}\otimes\tilde{U}(\Omega)$, we have
\begin{equation}
\label{zFECG_0_1_fun1}
U(\Omega) \bm{s}_J^{(r)} = \bm{u}_J^{(r)} \otimes\tilde{U}(\Omega)\bm{e}_z ~,
\end{equation}
where Eq.~(\ref{zFECG_0}) and the definition of $U(\Omega)$ in Eq.~(\ref{eq:RotOp}) have been exploited.

Considering Eqs.~(\ref{zFECG_0}), (\ref{zFECG_0_1_fun1}), and (\ref{zFECG_0_1_fun2}) 
and that $A_K^{(r)}=\bar{A}_K^{(r)}\otimes\mathbb{1}_3$ with $K\in\{I,J,IJ\}$, we have
\begin{equation}
\exp\left[2\, \bm{s}_I^{(r)^T} A_I^{(r)} A_{IJ}^{{(r)}^{-1}}
A_{J}^{(r)} U(\Omega) \bm{s}_J^{(r)} \right] 
= \exp\left[C\,\bm{e}_z^T\tilde{U}(\Omega)\bm{e}_z\right] 
= \exp\left[C\,\cos{\beta}\right] ~,
\label{zFECG_0_2}
\end{equation}
with $C$ given as
\begin{align}
C = 2\, \bm{u}_I^{(r)^T} \bar{A}_I^{(r)} \bar{A}_{IJ}^{{(r)}^{-1}}
\bar{A}_{J}^{(r)} \bm{u}_J^{(r)} ~.
\label{eq:zFECG_Cval}
\end{align}

Finally, the angular integration reduces to 
\begin{align}
\Upsilon^{N}_{M_N} = \frac{1}{4\pi^3} \int_0^{2\pi}d\alpha & \int_0^{\pi}d\beta \int_0^{2\pi}d\gamma 
\,\,\, \sin(\beta) \, D^{[N]^*}_{M_NM_N}(\Omega) \exp\left[C\cos(\beta)\right] ~,
\label{eq:UpsilonNMN}
\end{align}
To analytically solve the triple integration over Euler angles, we first note that 
the elements $D^{[N]}_{00}(\beta)$ of the Wigner $D$-matrices corresponding to $M_N=0$ are 
polynomial of $\cos\beta$ of degree $N$ with coefficients $a_\mu^{[N]}$ 
({\it{e.g.}}, $a_0^{[0]}=1$, $a_0^{[1]}=0$, $a_1^{[1]}=1$),
\begin{equation}
\label{eq:WignerN00}
D^{[N]}_{00}(\Omega) = D^{[N]}_{00}(\beta) = 
\sum_{\mu=0}^{N} a_\mu^{[N]} \left(\cos{\beta}\right)^{\mu} ~.
\end{equation}
Therefore, for apzFECGs with $M_N=0$, the integration over $\alpha$ and $\gamma$ Euler angles is trivial 
and Eq.~(\ref{eq:UpsilonNMN}) becomes
\begin{equation}
\Upsilon^{N}_{0} = \frac{1}{\pi} \sum_{\mu=0}^{N} \int_0^{\pi}d\beta 
\,\,\, \sin(\beta) [\cos(\beta)]^{\mu} \exp\left[C\cos(\beta)\right] ~.
\label{eq:UpsilonNMN_1}
\end{equation}
Furthermore, since apzFECG functions do not depend on Euler angles $\alpha$ and $\gamma$, 
the integration of the $D^{[N]^*}_{M_NM_N}(\Omega)$ yields zero for every $N\in\mathbb{N}_0$ and $M_N\ne0$.
The results of the integration over the Euler angle $\beta$ in Eq.~(\ref{eq:UpsilonNMN_1}) 
for the spherically symmetric ground state as well as the two lowest rotationally excited states are then written as
\small
\begin{align}
\Upsilon^{N}_{M_N} =
& \left\{ 
\begin{array}{lc} 
\displaystyle{\frac{2}{\pi C}} \sinh (C) 
& N=0 \,,\, M_N=0 \\[0.3cm]
\displaystyle{\frac{2}{\pi C}} \cosh (C) - \displaystyle{\frac{2}{\pi C^2}} \sinh (C) 
& N=1 \,,\, M_N=0 \\[0.3cm] 
\displaystyle{\frac{2}{\pi C^3}} \Big[\left(C^2+3\right) \sinh (C)-3 C \cosh (C)\Big]
& N=2 \,,\, M_N=0 \\[0.3cm] 
0 & \forall N\in\mathbb{N}_0 \,,\, M_N\ne0 
\end{array}
\right. ~,
\label{zFECG_0_3}
\end{align}
\normalsize
For a list of $\Upsilon^{N}_{M_N}$ up to $N=10$ see the Appendix.

\subsection{Kinetic integral}

The kinetic integral for plain FECGs reads \cite{Adamowicz2002_FECG}
\begin{align}
\left\langle \phi_I^{\text{FECG}} | -\bm{\nabla}_{\bm{r}}^TM\bm{\nabla}_{\bm{r}} | \phi_J^{\text{FECG}} \right\rangle
= \, \tilde{S}_{IJ} \, \Big[ 4\,\big( \bm{s}_I^{(r)}-\bm{s}_J^{(r)} \big)^T B 
\big(\bm{s}_I^{(r)}-\bm{s}_J^{(r)} \big) +6\Tr\left(M \bar{A}_J^{(r)} \bar{A}_{IJ}^{{(r)}^{-1}}
\bar{A}_I^{(r)} \right) \Big] ~,
\label{eq:kinetic_plainFECG}
\end{align}
where
\begin{align}
B = & \,\, 4 \, A_J^{(r)} A_{IJ}^{{(r)}^{-1}} A_I^{(r)} M
A_J^{(r)} A_{IJ}^{{(r)}^{-1}} A_I^{(r)} ~. 
\end{align}
For apzFECGs we have
\begin{equation}
T_{IJ[N,M_N,p]}^{\text{apzFECG}} = \left\langle \phi_I^{\text{zFECG}} \big| 
\hat{P}^{[N,p]}_{M_N} \phi_J^{\text{zFECG}} \right\rangle 
= \,\, \tilde{S}_{IJ} \, \Sigma^{N}_{M_N} ~,
\end{equation}
where the angular integral is written as 
\begin{align}
\Sigma^{N}_{M_N} =& \,\, \int \frac{d\Omega}{4\pi^3} \,\, D^{[N]}_{M_NM_N}(\Omega)^* \exp\left[ C\cos\beta \right] \nonumber \\[0.1cm]
& \times \left[ -\bm{s}_I^{(r)^T} B \bm{s}_I^{(r)}
-\bm{s}_J^{(r)^T} B \bm{s}_J^{(r)}
+2\bm{s}_I^{(r)^T} B\, U(\Omega) \bm{s}_J^{(r)}
+6\Tr\left(M \bar{A}_J^{(r)} \bar{A}_{IJ}^{{(r)}^{-1}} \bar{A}_I^{(r)} \right) \right] ~.
\label{zFECG_1}
\end{align}

We define 
\begin{equation}
\omega = -\bm{s}_I^{(r)^T} B \bm{s}_I^{(r)}
-\bm{s}_J^{(r)^T} B \bm{s}_J^{(r)}
+6\Tr\left(M \bar{A}_J^{(r)} \bar{A}_{IJ}^{{(r)}^{-1}} \bar{A}_I^{(r)} \right) ~,
\end{equation}
and 
\begin{equation}
\sigma= 2 \, \bm{u}_I^{(r)\,T} \bar{B} \bm{u}_J^{(r)} ~,
\end{equation}
so that Eq.~(\ref{zFECG_1}) can be cast in the compact form
\begin{align}
\Sigma^{N}_{M_N} =& \,\, 
\int \frac{d\Omega}{4\pi^3} \,\, D^{[N]}_{M_NM_N}(\Omega)^* \, \left( \omega+\sigma\cos\beta \right) \, \exp\left[ C\cos\beta \right] ~, 
\end{align}
With Eq.~(\ref{eq:WignerN00}), the integration over Euler angles can be reduced to the single integration over $\beta$ for which these
analytical results follow
\small
\begin{equation}
\Sigma^{N}_{M_N} =
\left\{ \begin{array}{lc} 
\displaystyle{\frac{2}{\pi C^2}} \Big[\sinh(C) (C\omega -\sigma)+C\sigma\cosh(C)\Big]  
& N=0 \,,\, M_N=0 \\[0.3cm]
\displaystyle{\frac{2}{\pi C^3}} \Big[\sinh(C) \left(\left(C^2+2\right) \sigma -C\omega \right)+C\cosh(C) (C\omega -2\sigma)\Big]  
& N=1 \,,\, M_N=0 \\[0.3cm] 
\displaystyle{\frac{2}{\pi C^4}} \Big[\sinh(C) \left(C \left(C^2+3\right) \omega -\left(4 C^2+9\right) \sigma\right) 
& \\[0.3cm]
\hspace{1cm} +C \cosh(C)\left(\left(C^2+9\right) \sigma -3C\omega \right)\Big] 
& N=2 \,,\, M_N=0 \\[0.3cm]
0 & \forall N\in\mathbb{N}_0 \,,\, M_N\ne0 \\
\end{array}\right. ~.
\end{equation}
\normalsize
For a list of $\Sigma^{N}_{M_N}$ up to $N=10$ see the Appendix.

\subsection{Coulomb integral}

From Ref.~\cite{Adamowicz2002_FECG} we retrieve the Coulomb matrix element for plain FECGs as follows:
\begin{align}
\left\langle \phi_I^{\text{FECG}} \left| \frac{1}{\left|\bm{r}_i-\bm{r}_j\right|} \right| \phi_J^{\text{FECG}} \right\rangle 
= \tilde{S}_{IJ} \, \left(\frac{1}{{\bm{S}}^TJ_{ij}{\bm{S}}}\right)^{\frac{1}{2}} 
\erf\left[\left(\frac{{\bm{S}}^TJ_{ij}{\bm{S}}}{\Tr\left(\bar{J}_{ij}
\bar{A}_{IJ}^{{(r)}^{-1}} \right)}\right)^{\frac{1}{2}}\right] ~,
\end{align} 
where the vector $\bm{S}$ is defined as
\begin{align}
{\bm{S}} = A_{IJ}^{{(r)}^{-1}} \left( A_I^{(r)}
\bm{s}_I^{(r)} +A_J^{(r)} \bm{s}_J^{(r)} \right) ~,
\end{align}
and
\begin{align}
J_{ij} = \left\{ \begin{array}{ll} E_{ii} & {\text{if}} \,\,\, i=j \\ E_{ii}+E_{jj}-E_{ij}-E_{ji} & {\text{if}} \,\,\, i\ne j \end{array} \right. ~,
\end{align}
with ${\left(E_{ij}\right)}_{\alpha\beta}=\delta_{\alpha\beta}$ being an $N_p\times N_p$ matrix.

We now define the matrix elements for apzFECG functions as 
\begin{align}
V_{IJ[N,M_N,p]}^{\text{apzFECG}} = \left\langle \phi_I^{\text{zFECG}} \left| \frac{1}{\left|\bm{r}_i-\bm{r}_j\right|}  
\right| \hat{P}^{[N,p]}_{M_N} \phi_J^{\text{zFECG}} \right\rangle 
= \,\, \tilde{S}_{IJ} \, \Lambda^{N}_{M_N} ~,
\end{align}
where
\begin{align}
\Lambda^{N}_{M_N} =& \,\, \int \frac{d\Omega}{4\pi^3} \,\, D^{[N]}_{M_NM_N}(\Omega)^* \, e^{C\cos\beta}
\left(\frac{1}{\tilde{\bm{S}}^TJ_{ij}\tilde{\bm{S}}}\right)^{\frac{1}{2}} 
\erf\left[\left(\frac{\tilde{\bm{S}}^TJ_{ij}\tilde{\bm{S}}}{\Tr\left(\bar{J}_{ij}
\bar{A}_{IJ}^{{(r)}^{-1}} \right)}\right)^{\frac{1}{2}}\right] ~.
\label{zFECG_2}
\end{align}

Here, we adopt the notation of Cafiero and Adamowicz \cite{Adamowicz2002_FECG} which is corrected in order to account 
for the rotated $\bm{s}_J$ vector
\begin{align}
\tilde{\bm{S}} = A_{IJ}^{{(r)}^{-1}} \left( A_I^{(r)} \bm{s}_I^{(r)}
+A_J^{(r)} U(\Omega)\bm{s}_J^{(r)} \right) ~.
\end{align}

In order to make $\beta$ explicit and solve the angular integration, we consider the following substitution
\begin{align}
\tilde{\bm{S}}^TJ_{ij}\tilde{\bm{S}} 
=& \tau_{ij} + 2\, \bm{s}_I^{(r)^T} A_I^{(r)} A_{IJ}^{{(r)}^{-1}}
J_{ij} A_{IJ}^{{(r)}^{-1}} A_J^{(r)} U(\Omega) \bm{s}_J^{(r)}
\nonumber \\[0.1cm]
=& \tau_{ij} + F_{ij} \left(\bm{e}_z^T\tilde{U}(\Omega)\bm{e}_z\right) 
\nonumber \\
=& \tau_{ij} + F_{ij} \, \cos\beta ~,
\end{align}
with
\begin{align}
\tau_{ij} = & \bm{s}_I^{(r)^T} A_I^{(r)} A_{IJ}^{{(r)}^{-1}}
J_{ij} A_{IJ}^{{(r)}^{-1}} A_I^{(r)} \bm{s}_I^{(r)}
+\bm{s}_J^{(r)^T} A_J^{(r)} A_{IJ}^{{(r)}^{-1}}
J_{ij} A_{IJ}^{{(r)}^{-1}} A_J^{(r)} \bm{s}_J^{(r)} ~, \\[0.1cm]
F_{ij} = & 2\cdot \bm{u}_I^{(r)^T} \bar{A}_I^{(r)}
\bar{A}_{IJ}^{{(r)}^{-1}} \bar{J}_{ij} \bar{A}_{IJ}^{{(r)}^{-1}}
\bar{A}_J^{(r)} \bm{u}_J^{(r)} ~.
\end{align}
The angular integration in Eq.~(\ref{zFECG_2}) is now written as
\begin{align}
\Lambda^{N}_{M_N} =& 
\int \frac{d\Omega}{4\pi^3} \,\, D^{[N]}_{M_NM_N}(\Omega)^* \, e^{C\cos\beta}
\left(\frac{1}{\tau_{ij} + F_{ij} \cdot \cos\beta}\right)^{\frac{1}{2}} 
\erf\left[\left(\frac{\tau_{ij} + F_{ij} \cdot \cos\beta}{\Tr\left(\bar{J}_{ij}
\bar{A}_{IJ}^{{(r)}^{-1}} \right)}\right)^{\frac{1}{2}}\right] ~.
\label{zFECG_2.1}
\end{align}
While the integration with respect to $\alpha$ and $\gamma$ is trivial to integrate over $\beta\in[0,\pi)$, we change the variable,
$y\equiv\tau_{ij}+F_{ij}\cos\beta$ so that Eq.~(\ref{zFECG_2.1}) becomes
\begin{align}
\Lambda^{N}_{M_N} =& \frac{e^{-\frac{\tau_{ij}\cdot C}{F_{ij}}}}{\pi F_{ij}} \int_{\tau_{ij}-F_{ij}}^{\tau_{ij}+F_{ij}} dy \,\, 
D^{[N]}_{M_NM_N}(y) \,\, y^{-\frac{1}{2}} \,\, e^{\frac{C}{F_{ij}}\,y} \,\,
\erf\left[\left(\frac{y}{\Tr\left(\bar{J}_{ij}
\bar{A}_{IJ}^{{(r)}^{-1}} \right)}\right)^{\frac{1}{2}}\right] ~.
\label{zFECG_2.2} 
\end{align}
To change the variable of the Wigner $D$-matrix we recall Eq.~(\ref{eq:WignerN00}), 
namely that the elements $D^{[N]}_{00}(\beta)$ for any $N$ are polynomial of $\cos\beta$ of degree $N$.
Therefore, after changing the variable, the zeroth diagonal element of the Wigner $D$-matrix can be written as
\begin{align}
D^{[N]}_{00}(y)
=& \sum_{\mu=0}^{N} a_{\mu}^{[N]} \left( \frac{y-\tau_{ij}}{F_{ij}} \right)^{\mu} \nonumber \\[0.1cm]
=& \sum_{\mu=0}^{N} \sum_{k=0}^{\mu} \frac{\mu!\,a_\mu^{[N]}}{(\mu-k)!k!}
\label{zFECG_Dpoly}
\end{align}
where in the second line the power of the binomial is written explicitly.
By inserting Eq.~(\ref{zFECG_Dpoly}), the polynomial form of the Wigner $D$-matrix, 
Eq.~(\ref{zFECG_2.2}) reads
\begin{align}
\Lambda^{N}_{0} =& \,\, \frac{e^{-\frac{\tau_{ij}\cdot C}{F_{ij}}}}{\pi F_{ij}} 
\sum_{\mu=0}^{N} \sum_{k=0}^{\mu} \frac{\mu!\,a_\mu^{[N]}}{(\mu-k)!k!}  
\left(-\frac{\tau_{ij}}{F_{ij}}\right)^{\mu-k} \left(\frac{1}{F_{ij}}\right)^k \nonumber \\[0.1cm]
& \times \int_{\tau_{ij}-F_{ij}}^{\tau_{ij}+F_{ij}} dy \,\, y^{-\frac{1}{2}+k} \,\, 
e^{\frac{C}{F_{ij}}\,y} \,\, \erf\left[\left(\frac{y}{\Tr\left(\bar{J}_{ij}
\bar{A}_{IJ}^{{(r)}^{-1}} \right)}\right)^{\frac{1}{2}}\right] ,
\label{zFECG_2.21} 
\end{align}
whereas expanding the exponential in a Taylor series yields
\begin{align}
\Lambda^{N}_{0} =& \,\, \frac{e^{-\frac{\tau_{ij}\cdot C}{F_{ij}}}}{\pi F_{ij}} 
\sum_{\mu=0}^{N} \sum_{k=0}^{\mu} \frac{\mu!\,a_\mu^{[N]}}{(\mu-k)!k!}  
\left(-\frac{\tau_{ij}}{F_{ij}}\right)^{\mu-k} \left(\frac{1}{F_{ij}}\right)^k \nonumber \\[0.1cm]
& \times \sum_{n=0}^{\infty} \frac{1}{n!} \left(\frac{C}{F_{ij}}\right)^{n} 
\int_{\tau_{ij}-F_{ij}}^{\tau_{ij}+F_{ij}} dy \,\,
\,\, y^{-\frac{1}{2}+k+n} \,\, \erf\left[\left(\frac{y}{\Tr\left(\bar{J}_{ij}
\bar{A}_{IJ}^{{(r)}^{-1}} \right)}\right)^{\frac{1}{2}}\right] ~.
\label{zFECG_2.22}
\end{align}
The integral over $y$ possesses an analytical solution,
\begin{align}
\Lambda^{N}_{0} =& \frac{e^{-\frac{\tau_{ij}\cdot C}{F_{ij}}}}{\pi F_{ij}} 
\sum_{\mu=0}^{N} \sum_{k=0}^{\mu} \frac{\mu!\,a_\mu^{[N]}}{(\mu-k)!k!}  
\left(-\frac{\tau_{ij}}{F_{ij}}\right)^{\mu-k} \left(\frac{1}{F_{ij}}\right)^k 
\sum_{n=0}^{\infty} \frac{2}{(2 k+2 n+1) n!} \left(\frac{C}{F_{ij}}\right)^{n} \nonumber \\[0.1cm]
& \, \times \Bigg[ -\erf(\sqrt{t_2}) \, (\tau_{ij}-F_{ij})^{k+n+\frac{1}{2}} 
+\erf(\sqrt{t_1}) \, (F_{ij}+\tau_{ij})^{k+n+\frac{1}{2}} \nonumber \\[0.1cm]
& \hspace{0.6cm} + \frac{\Tr\big(\bar{J}_{ij}
\bar{A}_{IJ}^{{(r)}^{-1}} \big)^{k+n+\frac{1}{2}}}{\sqrt{\pi}} \,\, 
\bigg( \Gamma (k+n+1,t_1) \,-\, \Gamma (k+n+1,t_2) \bigg) \Bigg] ~,
\label{zFECG_2.3}
\end{align}
with 
\begin{align}
t_1 = & \frac{\tau_{ij}+F_{ij}}{\Tr\big(\bar{J}_{ij}\bar{A}_{IJ}^{{(r)}^{-1}}\big)} ~, \\[0.1cm]
t_2 = & \frac{\tau_{ij}-F_{ij}}{\Tr\big(\bar{J}_{ij}\bar{A}_{IJ}^{{(r)}^{-1}}\big)} ~.
\end{align}
If the resulting series in Eq.~(\ref{zFECG_2.3}) is considered separately for each term, 
the first two can be evaluated exactly in terms of the lower incomplete Gamma function $\gamma(n,b)$, 
while the latter is simplified according to the properties of the incomplete Gamma functions 
\begin{align}
\Lambda^{N}_{0} =& \frac{e^{-\frac{\tau_{ij}\cdot C}{F_{ij}}}}{\pi F_{ij}} 
\sum_{\mu=0}^{N} \sum_{k=0}^{\mu} \,\, \frac{\mu!\,a_\mu^{[N]}}{(\mu-k)!k!} \,
\left(-\frac{\tau_{ij}}{F_{ij}}\right)^{\mu-k} \, \left(\frac{1}{F_{ij}}\right)^k \nonumber \\[0.1cm]
& \Bigg[ \left(-\frac{C}{F_{ij}}\right)^{-k-\frac{1}{2}} \erf(\sqrt{t_1}) 
\,\,\, \gamma\left(k+\frac{1}{2},-\frac{C(F_{ij}+\tau_{ij})}{F_{ij}}\right) \nonumber \\[0.1cm]
& -\left(-\frac{C}{F_{ij}}\right)^{-k-\frac{1}{2}} \erf(\sqrt{t_2}) 
\,\,\, \gamma\left(k+\frac{1}{2},\frac{C(F_{ij}-\tau_{ij})}{F_{ij}}\right) \nonumber \\[0.1cm]
& + \frac{2}{\sqrt{\pi}} \sum_{n=0}^{\infty} \frac{\Gamma (k+n+1,t_1,t_2)}{n! \,\, (2 k+2 n+1)}
\left(\frac{C}{F_{ij}}\right)^n \Tr\big(\bar{J}_{ij}
\bar{A}_{IJ}^{{(r)}^{-1}} \big)^{k+n+\frac{1}{2}} \Bigg] ~,
\label{zFECG_2.4}
\end{align}
where the last remaining series converges factorially and only requires the generalized incomplete Gamma functions
$\Gamma(n,a,b)$, with $n\in\mathbb{N}^+$, that can be efficiently calculated in closed form as
\begin{align}
\Gamma(n,t_1,t_2) = \Gamma(n) \left( e^{-t_1} \sum_{k=0}^{n-1} \frac{t_1^k}{k!} 
-e^{t_2} \sum_{k=0}^{n-1} \frac{t_2^k}{k!} \right) ~.
\end{align}
While Eq.~(\ref{zFECG_2.4}) provides a general $N$-formula toward the calculation of Coulomb matrix elements, 
a closed formula can be obtained with the 'differentiation under the integral' technique 
from Eq.~(\ref{zFECG_2.21})
\begin{align}
\Lambda^{N}_{0} =& \frac{e^{-\frac{\tau_{ij}\cdot C}{F_{ij}}}}{\pi F_{ij}} 
\sum_{\mu=0}^{N} \sum_{k=0}^{\mu} \,\, \frac{\mu!\,a_\mu^{[N]}}{(\mu-k)!k!} \,
\left(-\frac{\tau_{ij}}{F_{ij}}\right)^{\mu-k} \, \left(\frac{1}{F_{ij}}\right)^k \nonumber \\[0.2cm]
& \times 2 \, T^{\frac{1}{2}}\,F_{ij}^k \,\, \frac{\partial^k}{\partial C^k} 
\int_{\sqrt{(\tau_{ij}-F_{ij})/T}}^{\sqrt{(\tau_{ij}+F_{ij})/T}} 
dx \,\, e^{\frac{C\,T}{F_{ij}}\,x^2} \,\, \erf\left[x\right] ~,
\label{zFECG_2.5} 
\end{align}
where $T=\Tr(\bar{J}_{ij} \bar{A}_{IJ}^{{(r)}^{-1}})$, 
the integration variable is changed according to $(y/T)^{\frac{1}{2}}=x$,
and the $k$-th derivative with respect to $C$ is considered. 
The integral in Eq.~(\ref{zFECG_2.5}) possesses an analytical solution,
\begin{align}
& \int_a^b dx \,\, e^{-q\,x^2} \,\, \erf\left[x\right] = \, 2\sqrt{\frac{\pi}{q}} 
\Bigg[ T\Big(a\sqrt{2q},\frac{1}{\sqrt{q}}\Big) - T\Big(b\sqrt{2q},\frac{1}{\sqrt{q}}\Big) \Bigg] ~,
\label{zFECG_2.6}
\end{align}
where $T(h,x)$ is the Owen's T function.

\subsection{Squared total angular momentum expectation value}

To solve $\langle\phi_I^{\text{zFECG}}|\hat{N}^2|\phi_J^{\text{apzFECG}}\rangle$,
the squared total angular momentum expectation value for projected zFECG functions,
we start from the matrix elements for FECGs derived in our previous work \cite{Muolo2018b}
\begin{align}
\langle\phi_I^{\text{FECG}}|\hat{N}^2|\phi_J^{\text{FECG}}\rangle =& \epsilon_{ijk}'  
\Bigg[ 2 \bigg( \bm{s}_I^{(r)^T} \omega_I^{(j,k)^T} 
A_{IJ}^{{(r)}^{-1}} \omega_{J}^{(j,k)} \bm{s}_J^{(r)} \bigg) 
\nonumber \\[0.2cm]
& + 4 \bigg( {\bm{w}}^T A_{IJ}^{{(r)}^{-1}} \omega_J^{(j,k)} \bm{s}_J^{(r)} \bigg) 
\bigg( \bm{w}^T A_{IJ}^{{(r)}^{-1}} \omega_{I}^{(j,k)} \bm{s}_I^{(r)} \bigg) \Bigg] 
S_{IJ}^{\text{FECG}} ~,
\end{align}
where $\bm{w}=A_I^{(r)}\bm{s}_I^{(r)}+A_J^{(r)}\bm{s}_J^{(r)}$, 
$\epsilon_{ijk}'$ is the Levi-Civita symbol for which only the negative entries are set to zero,
and 
\begin{align}
\label{zFECG_4}
\omega^{(x,y)}_K = \bar{A}_K^{(r)} \otimes 
\left( E_{xy}-E_{yx} \right) \quad {\text{with}} \,\,\, K\in\{I,J\} ~,
\end{align}
with $(E_{ij})_{xy}=\delta_{ix}\delta_{jy}$.
Note that the $i,j,$ and $k$ indices are summed with Einstein's summation convention.
We recall that for {apzFECG} functions, the vector $\bm{s}_K^{(r)}$ ($K\in\{I,J\}$) must obey the constraint introduced in Eq.~(\ref{zFECG_0})
and $\bm{s}_J^{(r)}$ is subject to the rotation operator $\hat{R}(\Omega)$ involving 
the transformation matrix $U(\Omega)$. 
Considering Eqs.~(\ref{zFECG_0}), (\ref{zFECG_0_1_fun1}), (\ref{zFECG_0_1_fun2}) and (\ref{zFECG_4}) we have
\begin{align}
\langle\phi_I^{\text{zFECG}}|\hat{N}^2|\phi_{J[N,M_N,p]}^{\text{apzFECG}}\rangle = & \, \tilde{S}_{IJ} \,\, \Xi^{N}_{M_N} ~,
\end{align}
where 
\begin{align}
\Xi^{N}_{M_N} = & \, \epsilon_{ijk}' \int \frac{d\Omega}{4\pi^3} \,\, D_{M_NM_N}^{[N]}(\Omega)^* \, e^{C\,\cos\beta} \nonumber \\[0.1cm]
& \Bigg[ 2 \big( \bm{u}_I^{(r)^T} \bar{A}_I^{(r)} \bar{A}_{IJ}^{{(r)}^{-1}}
\bar{A}_J^{(r)} \bm{u}_J^{(r)} \big) 
\Big(\bm{e}_z^T(E_{jk}-E_{kj})^T(E_{jk}-E_{kj})\tilde{U}(\Omega)\bm{e}_z\Big) \nonumber \\[0.1cm]
& +4 \bigg( \big( \bm{u}_I^{(r)^T} \bar{A}_I^{(r)} \bar{A}_{IJ}^{{(r)}^{-1}}
\bar{A}_J^{(r)} \bm{u}_J^{(r)} \big) 
\big(\bm{e}_z^T(E_{jk}-E_{kj})\tilde{U}(\Omega)\bm{e}_z\big) \nonumber \\[0.1cm]
& + \big(\bm{u}_J^{(r)^T} \bar{A}_J^{(r)} \bar{A}_{IJ}^{{(r)}^{-1}}
\bar{A}_J^{(r)} \bm{u}_J^{(r)} \big)  
\big(\bm{e}_z^T\tilde{U}(\Omega)^T(E_{jk}-E_{kj})\tilde{U}(\Omega)\bm{e}_z\big) \bigg) \nonumber \\[0.1cm]
& \times \bigg( \big( \bm{u}_I^{(r)^T} \bar{A}_I^{(r)} \bar{A}_{IJ}^{{(r)}^{-1}}
\bar{A}_I^{(r)} \bm{u}_I^{(r)} \big) 
\big(\bm{e}_z^T(E_{jk}-E_{kj})\bm{e}_z\big) \nonumber \\[0.1cm]
& + \big(\bm{u}_J^{(r)^T} \bar{A}_J^{(r)} \bar{A}_{IJ}^{{(r)}^{-1}}
\bar{A}_I^{(r)} \bm{u}_I^{(r)} \big) 
\big(\bm{e}_z^T\tilde{U}(\Omega)^T(E_{jk}-E_{kj})\bm{e}_z\big) \bigg) \Bigg] ,
\label{zFECG_4_1}
\end{align}
where $C$ has been defined in Eq.~(\ref{eq:zFECG_Cval}).

Furthermore, provided that $(j,k)\in\{(2,3),(3,1),(1,2)\}$ (see Ref.~\cite{Muolo2018b} for a detailed demonstration), we have
\begin{align}
& \bm{e}_z^T(E_{23}-E_{32})^T(E_{23}-E_{32})\tilde{U}(\Omega)\bm{e}_z = \cos\beta \label{eq:zFECG_transfProp1} ~, \\[0.1cm]
& \bm{e}_z^T(E_{31}-E_{13})^T(E_{31}-E_{13})\tilde{U}(\Omega)\bm{e}_z = \cos\beta \label{eq:zFECG_transfProp2} ~, \\[0.1cm]
& \bm{e}_z^T(E_{23}-E_{32})\tilde{U}(\Omega)\bm{e}_z = +\sin\alpha \, \sin\beta \label{eq:zFECG_transfProp3} ~, \\[0.1cm]
& \bm{e}_z^T(E_{31}-E_{13})\tilde{U}(\Omega)\bm{e}_z = -\cos\alpha \, \sin\beta \label{eq:zFECG_transfProp4} ~, \\[0.1cm]
& \bm{e}_z^T\tilde{U}(\Omega)^T(E_{23}-E_{32})\tilde{U}(\Omega)\bm{e}_z = 0 \label{eq:zFECG_transfProp5} ~, \\[0.1cm]
& \bm{e}_z^T\tilde{U}(\Omega)^T(E_{31}-E_{13})\tilde{U}(\Omega)\bm{e}_z = 0 \label{eq:zFECG_transfProp6} ~, \\[0.1cm]
& \bm{e}_z^T(E_{23}-E_{32})\bm{e}_z = 0 \label{eq:zFECG_transfProp7} ~, \\[0.1cm]
& \bm{e}_z^T(E_{31}-E_{13})\bm{e}_z = 0 \label{eq:zFECG_transfProp8} ~, \\[0.1cm]
& \bm{e}_z^T\tilde{U}(\Omega)^T(E_{23}-E_{32})\bm{e}_z = -\sin\alpha \, \sin\beta \label{eq:zFECG_transfProp9} ~, \\[0.1cm]
& \bm{e}_z^T\tilde{U}(\Omega)^T(E_{31}-E_{13})\bm{e}_z = +\cos\alpha \, \sin\beta \label{eq:zFECG_transfProp10} ~,
\end{align}
while it can be shown that for $(j,k)=(1,2)$ all these expressions evaluate to zero.

Eq.~(\ref{zFECG_4_1}) can now be written as
\begin{align}
\Xi^{N}_{M_N} =& \int \frac{d\Omega}{4\pi^3} \,\, D_{M_NM_N}^{[N]}(\Omega)^* \, 
\exp\big[{C\cos\beta}\big]
\nonumber \\[0.2cm]
& \times \Bigg[ 2 \big( C\,\cos\beta \big) + \big(C\sin\alpha\,\sin\beta\big) \big(-C\sin\alpha\,\sin\beta\big) \nonumber \\
& + \big(-C\cos\alpha\,\sin\beta\big) \big(C\cos\alpha\,\sin\beta\big) \Bigg] ~,
\end{align}
and its analytical solution to the angular integration for $N=0,1,$ and $2$ yields 
\small
\begin{align}
\Xi^{N}_{M_N} =&
\left\{ \begin{array}{lc} 
0                    & \quad {\text{if}} \quad N=0 \,,\, M_N=0 \\ 
2 \,\, \Upsilon^1_0  & \quad {\text{if}} \quad N=1 \,,\, M_N=0 \\ 
6 \,\, \Upsilon^2_0  & \quad {\text{if}} \quad N=2 \,,\, M_N=0 \\ 
0                    & \quad \forall \, N\in\mathbb{N}_0   \,,\, M_N\ne0 
\end{array} \right. ~,
\end{align}
\normalsize
where $\Upsilon^{N}_{M_N}$ are the solution of the overlap angular integration given in Eq.~(\ref{zFECG_0_3}).
This is in accordance with the expected eigenvalue for the squared total spatial angular momentum $N(N+1)$ in Hartree atomic units.
For a list of $\Xi^{N}_{M_N}$ up to $N=5$ see the Appendix.

\subsection{Projection of the angular momentum onto the $z$ axis}

We recall the $\langle\hat{N}_z\rangle_{IJ}$ matrix elements for FECG functions \cite{Muolo2018b}
\begin{align}
\langle\phi_I^{\text{FECG}}|\hat{N}_z|\phi_J^{\text{FECG}}\rangle = 
\frac{2}{i} \left( \bm{w}^T A_{IJ}^{{(r)}^{-1}} \omega_J^{(1,2)} \bm{s}_J^{(r)} \right)
\left\langle\phi_I|\phi_j\right\rangle ~.
\end{align}
Here, we cannot simplify the expectation value for apzFECGs since $[\hat{R}(\Omega),\hat{N}_z]\ne0$.
The term in parenthesis then becomes  
\begin{align}
\bm{w}^T A_{IJ}^{{(r)}^{-1}} \omega_J^{(1,2)} \bm{s}_J^{(r)}
=& \bm{s}_I^{(r)^T} A_I^{(r)} A_{IJ}^{{(r)}^{-1}}
\omega_J^{(1,2)} \bm{s}_J^{(r)}
+ \bm{s}_J^{(r)^T} A_J^{(r)}
A_{IJ}^{{(r)}^{-1}} \omega_J^{(1,2)} \bm{s}_J^{(r)} \nonumber \\[0.2cm]
=& \left( \bm{u}_I^{(r)^T} \bar{A}_I^{(r)} \bar{A}_{IJ}^{{(r)}^{-1}}
\bar{A}_J^{(r)} \bm{u}_J^{(r)} \right) 
\left(\bm{e}_z^T\tilde{U}(\Omega')(E_{21}-E_{12})\tilde{U}(\Omega)\bm{e}_z\right) \nonumber \\[0.2cm]
& + \left( \bm{u}_J^{(r)^T} \bar{A}_J^{(r)} \bar{A}_{IJ}^{{(r)}^{-1}}
\bar{A}_J^{(r)} \bm{u}_J^{(r)} \right) 
\left(\bm{e}_z^T\tilde{U}(\Omega)^T(E_{21}-E_{12})\tilde{U}(\Omega)\bm{e}_z\right) =0 ~.
\end{align}
It follows from Eqs.~(\ref{eq:zFECG_transfProp5}) and (\ref{eq:zFECG_transfProp6}) that the latter term is zero, 
i.e., $\bm{e}_z^T(E_{21}-E_{12})\tilde{U}(\Omega)\bm{e}_z=0$, 
while the former one is
\begin{align}
\bm{e}_z^T\tilde{U}(\Omega')(E_{21}-E_{12})\tilde{U}(\Omega)\bm{e}_z = 
\cos\alpha'\sin\alpha\sin\beta\sin\beta' - \cos\alpha\sin\alpha'\sin\beta\sin\beta' ~.
\end{align}
The resulting expectation value for apzFECG functions reads
\begin{align}
\langle\phi_{I[N,M_N,p]}^{\text{apzFECG}}|\hat{N}_z|\phi_{J[N,M_N,p]}^{\text{apzFECG}}\rangle =& 
\, \tilde{S}_{IJ} \int\frac{d\Omega}{4\pi^3} \int\frac{d\Omega'}{4\pi^3} \,\, D_{M_NM_N}^{[N]}(\Omega)^* D_{M_NM_N}^{[N]}(\Omega')^* 
\exp\left[C\,\cos\beta\right] \nonumber \\[0.2cm]
& \times \left( \bm{u}_I^{(r)^T} \bar{A}_I^{(r)} \bar{A}_{IJ}^{{(r)}^{-1}}
\bar{A}_J^{(r)} \bm{u}_J^{(r)} \right) 
\left[\sin(\alpha-\alpha')(\sin\beta)^2(\sin\beta')^2\right] ~, 
\end{align}
which evaluates to zero for every $N$, $M_N$ pairs:
\begin{align}
\langle\phi_{I[N,M_N,p]}^{\text{apzFECG}}|\hat{N}_z|\phi_{J[N,M_N,p]}^{\text{apzFECG}}\rangle 
= 0 \quad \forall \,\, N \,\, | \,\, N=(0,1,2,\ldots) , M_N=(-N,\ldots,+N) ~.
\end{align}

This shows that apzFECG functions have zero projection of the total angular momentum on the $z$ axis.
The results in this section can be expanded by noting that not only the expectation value 
of $\hat{N}_z$ is zero, but also the corresponding eigenvalue of the apzFECG functions, 
\begin{align}
\hat{N}_z \phi_{I\,[N,M_N,p]}^{\text{apzFECG}} = 0 ~.
\label{eq:apzFECG_eigenfun_Nz}
\end{align}
The derivation of Eq.~(\ref{eq:apzFECG_eigenfun_Nz}) follows from the definition of 
$\phi_{I\,[N,M_N,p]}^{\text{apzFECG}}$, $\hat{N}_z$, and $P^{[N,p]}_{M_N}$
\begin{align}
\hat{N}_z \phi_{I\,[N,M_N,p]}^{\text{apzFECG}} =& \hat{N}_z P^{[N,p]}_{M_N} 
\phi_I^{\text{zFECG}} \big(\bm{r};A_I^{(r)},\bm{s}_I^{(r)}\big)
\nonumber \\[0.1cm]
=& \hat{N}_z \int \frac{d\Omega}{4\pi^3} ~ D_{M_NM_N}^{[N]}(\Omega)^* 
\phi_I^{\text{zFECG}} \Big(\bm{r};A_I^{(r)},U(\Omega)\bm{s}_I^{(r)}\Big)
\nonumber \\[0.1cm]
=& \frac{2}{i} \int \frac{d\Omega}{4\pi^3} ~ D_{M_NM_N}^{[N]}(\Omega)^* 
\Big[\bm{r}^T\omega_I^{(x,y)}U(\Omega)\bm{s}_I^{(r)}\Big]
\phi_I^{\text{zFECG}} \Big(\bm{r};A_I^{(r)},U(\Omega)\bm{s}_I^{(r)}\Big) ~,
\label{eq:apzFECG_eigenfun_Nz_1}
\end{align}
and by noting that 
\begin{align}
U(\Omega)\bm{s}_I^{(r)} = \bm{u}_I^{(r)} \otimes \tilde{U}(\Omega)\bm{e}_z
= \bm{u}_I^{(r)} \otimes \left(
\begin{array}{c} -\cos{\alpha}\sin{\beta} \\ -\sin{\alpha}\sin{\beta}\cos{\beta} \\ \cos{\beta} \end{array}
\right) ~.
\end{align}
Since $D_{M_NM_N}^{[N]}\propto\exp(-\text{i}M_N\gamma)$ and the right-hand side 
of Eq.~(\ref{eq:apzFECG_eigenfun_Nz_1}) do not depend on $\gamma$,
the integration over the Euler angles yields zero for all $M_N\ne0$.
This shows that Eq.~(\ref{eq:apzFECG_eigenfun_Nz}) is correct, and additionally, we have 
\begin{align}
\phi_{I\,[N,M_N,p]}^{\text{apzFECG}} 
= \hat{P}_{M_N}^{[N,p]} \phi_I^{\text{zFECG}}   
= 0 ~~~ \forall ~ M_N\ne0 ~,
\end{align}
i.e., there is no component of $\phi_I^{\text{zFECG}}$ on the $M_N\ne0$ eigenspaces.

\subsection{Elimination of center-of-mass contamination}

Contributions from the center of mass are eliminated from the expectation values according 
to the protocol devised in Refs.~\cite{Benjamin2013,Muolo2018a}. 
First, the variational matrices $A^{(r)}$ and the variational vectors $\bm{s}^{(r)}$ are 
manipulated in a given TICC, 
$A^{(x)}$ and $\bm{s}^{(x)}$, respectively, and defined in block diagonal form
\begin{align}
\bar{A}_I^{(r)} =& \, U_x^{T} \left( 
\begin{array}{cc} \mathcal{A}_I^{(x)} & 0 \\ 0 & {c_A} \end{array} \right) U_x ~, \\[0.2cm]
\bm{s}_I^{(r)} =& U_x \left(\begin{array}{cc} \bm{s}_I^{(x)} \\ {\bm{c}_S} \end{array} \right) 
= U_x  \left(\begin{array}{cc} \bm{u}_I^{(x)} \\ {c_S}_z \end{array}\right) 
\otimes \bm{e}_z ~,
\end{align}
where the $N_p-1\times N_p-1$ matrix $\mathcal{A}_I^{(x)}$
and the $N_p-1$ vector $\bm{u}_I^{(x)}$ are related to the internal coordinates, 
while $c_A$ and ${c_S}_z$ are scalar parameters associated with the center of mass. 
Note the superscript distinguishing the LFCC set $\{r\}$ from a generic TICC set $\{x\}$.
Although the choice of zero for both $c_A$ and ${c_S}_z$ for all $I\in\{1,\ldots,N_b\}$ 
would systematically cancel center-of-mass contributions from every expectation value, 
$c_A=0$ leads to a singular matrix $A_I$, which violates the square-integrable 
and positive-definiteness requirements for the basis functions.

We note that the choice of $c_{A}=1$ and ${c_S}_z=0$, 
implies that every FECG, zFECG, or apzFECG function is exactly factorizable into 
a spherical Gaussian centered at the origin for the center-of-mass coordinate, 
and an FECG function for the $N_p-1$ internal coordinates. 
In fact, the FECG in (transformed) TICC coordinates $\{x\}$ can be written as
\begin{align}
\phi_I^{\text{FECG}} =& \exp\Bigg[ -
\left(\begin{array}{cc} \bm{x}-\bm{s}_I^{(x)} \\ \bm{x}_{\text{CM}}-{\bm{c}_S} \end{array} \right)^T
\left( \begin{array}{cc} \mathcal{A}_I^{(x)} & 0 \\ 0 & {c_A} \end{array} \right)
\left(\begin{array}{cc} \bm{x}-\bm{s}_I^{(x)} \\ \bm{x}_{\text{CM}}-{\bm{c}_S} \end{array} \right)
\Bigg] \nonumber \\[0.2cm]
=& \exp\left[ -(\bm{x}-\bm{s}_I^{(x)})^T \mathcal{A}_I^{(x)} (\bm{x}-\bm{s}_I^{(x)}) \right]
\exp\Big[-\bm{x}_{\text{CM}}^2\Big] ~.
\end{align}

We chose not to evaluate the integral matrix elements with basis functions and operators 
in a (transformed) TICC set. Instead, we carry out the integrations straightforwardly
in the simple LFCC set and correct \textit{a posteriori} 
the resulting expression by subtracting center-of-mass dependent terms
as described in our previous work. 
Hence, elimination of center-of-mass contaminations is equivalent to subtraction of the 
residual $c_A$-terms \cite{Benjamin2013,Muolo2018a}.

We start detecting $c_{A}$-dependent terms from the $C$ factor.
To this aim, we transform it to the TICC sets $\{x\}$ and $\{y\}$, for the 
$I$-th and $J$-th basis functions, respectively,
\begin{align}
C =& \,2\,\bm{u}_I^{(r)\,T}\bar{A}_I^{(r)}\bar{A}_{IJ}^{{(r)}^{-1}}\bar{A}_{J}^{(r)}\bm{u}_J^{(r)} 
  \nonumber \\[0.2cm]
  =& \,2\,\bm{u}_I^{(r)\,T} \Big[ U_x^{T}\bar{A}_I^{(x)}U_x\bar{A}_{IJ}^{{(r)}^{-1}}
  U_y^{T}\bar{A}_I^{(y)}U_y \Big] \bm{u}_J^{(r)}
  \nonumber \\[0.2cm]
  =& \,2\,\bm{u}_I^{(r)\,T} \left[ U_x^T
  \left(\begin{array}{cc} \mathcal{A}_{I}^{(x)} & 0 \\ 0 & {c_A} \end{array}\right) 
  \left(\begin{array}{cc} \mathcal{A}_{IJ}^{-1} & 0 \\ 0 & \frac{1}{2c_A} \end{array}\right)
  \left(\begin{array}{cc} \mathcal{A}_{J}^{(y)} & 0 \\ 0 & {c_A} \end{array}\right)
  U_x \right] \bm{u}_J^{(r)}
  \nonumber \\[0.2cm]
  =& \,2\, \big( \bm{u}_I^{(x)} ~~ {c_S}_z \big)
  \left(\begin{array}{cc} \mathcal{A}_{I}^{(x)}\mathcal{A}_{IJ}^{-1}\mathcal{A}_{J}^{(y)}
  & 0 \\ 0 & \frac{c_A}{2} \end{array}\right) 
  \left(\begin{array}{c} \bm{u}_J^{(y)} \\ {c_S}_z \end{array}\right) ~,
  \label{eq:C_TICC}
\end{align}
where $\mathcal{A}_{IJ}=\mathcal{A}_I+\mathcal{A}_J$. In the third step,
the following mathematical relation is employed \cite{Muolo2018a}
\begin{align}
  U_x\bar{A}_{IJ}^{-1}U_y^T = 
  \left(\begin{array}{cc} \mathcal{A}_{IJ}^{-1} & 0 \\ 0 & \frac{1}{2c_A} \end{array}\right) ~.
\end{align}
From Eq.~(\ref{eq:C_TICC}) it follows that the center-of-mass contributions to $C$ are zero for ${c_S}_z=0$.
For this reason, since the expectation value of the total angular momentum squared operator 
depends solely on $C$ terms, we conclude that it is free of center-of-mass contaminations.

The only center-of-mass dependent term arising in the analytical kinetic energy integral 
with the favorable choice ${c_S}_z=0$, is the $R$ term defined as 
\begin{align}
R = \Tr\left(M A_J^{(r)} A_{IJ}^{{(r)}^{-1}} A_I^{(r)}\right) ~.
\end{align}
The translational contamination can now be eliminated by replacing
\begin{align}
R_{\text{corr.}} = R - \frac{1}{4} {c_A}{c_M} ~,
\end{align}
with $c_M=\sum_{i=0}^{N_p}m_i$ being the total mass of the system.
We emphasize that minimization of the energy with respect to translationally invariant parameters only 
excludes the center-of-mass coordinate, and hence, reduces the original problem for $N_p$ particles to a simpler optimization problem 
for $N_p-1$ pseudo-particles with lower complexity.

\subsection{Numerical stability} \label{SEC:NumStab}

We investigate the numerical stability of the analytical matrix elements in finite-precision arithmetic. 
A naive implementation of the integral expressions results in ill-conditioned overlap and Hamiltonian matrices 
because of the hyperbolic functions. 
To restore numerical stability, we introduce normalization for the basis functions, defined as
\begin{equation}
\Phi_{I\,[N,M_N,p]}^{\text{apzFECG}} = \frac{\hat{P}^{[N,p]}_{M_N}\phi^{\text{zFECG}}_I}{|\phi_{I\,[N,M_N,p]}^{\text{apzFECG}}|} ~,
\end{equation}
where the normalization factor is
\begin{align}
|\phi^{[N,M_N]}_I|=\langle
\hat{P}^{[N,p]}_{M_N}\phi^{\text{apzFECG}}_{I[N,M_N,p]}|
\hat{P}^{[N,p]}_{M_N}\phi^{\text{apzFECG}}_{I[N,M_N,p]}\rangle^{\frac{1}{2}} ~.
\end{align}
Matrix elements $\mathcal{O}_{IJ}^{\text{apzFECG}}$ for a generic operator $\hat{O}$ are then evaluated as
\begin{equation}
\langle\Phi^{\text{apzFECG}}_{I[N,M_N,p]}|\hat{O}|\Phi^{\text{apzFECG}}_{J[N,M_N,p]}\rangle = 
\frac{\langle\hat{P}^{[N,p]}_{M_N}\phi^{\text{zFECG}}_{I[N,M_N,p]} |\hat{O}| \hat{P}^{[N,p]}_{M_N}\phi^{\text{zFECG}}_{J[N,M_N,p]}\rangle}
{|\phi^{\text{apzFECG}}_{I[N,M_N,p]}| |\phi^{\text{apzFECG}}_{J[N,M_N,p]}|} ~.
\end{equation}
Although the normalization of apzFECGs assures well-conditioned representation matrices for the quantum mechanical operators,
extreme $C$ values cause overflow of the hyperbolic sine and cosine functions
as well as cancellation errors in the kinetic energy terms because of the high powers of $C$. 
To remedy these two sources of errors, we differentiate the integral evaluation scheme for different orders of magnitude of $C$
by allowing higher-precision arithmetic to be employed when needed. 
In particular, we detected possible sources of numerical instabilities for $|C|>700$ when working 
in double precision floating point arithmetic. However,
quadruple precision suffices for achieving the desired accuracy for every test calculations 
with unconstrained optimization of the variational parameters.
While basis functions yielding $|C|>700$ can also be discarded,
we prefer the latter strategy to keep the energy function continuous with respect to the variational parameters.

The accuracy and convergence of special functions, 
i.e., the hyperbolic sine and cosine functions and the generalized incomplete Gamma functions, 
converge to $0.9~\varepsilon$ for every point without the need to resort to higher precision arithmetics.
The latter we implemented for the handling of particularly difficult cases following 
Ref.~\cite{GenIncompGamma_1994,GenIncompGamma_1996}.

Comparing apzFECGs for $N=0$ and the spherically symmetric (simple) ECG functions,
we note that the former require systematically less function evaluations to reach a given accuracy.
Simple ECG functions are plagued by problems of linear dependence in the basis 
during energy optimization of a polyatomic system.
In diatomics, there exists a large nuclear density at a distance to the origin in relative coordinates. 
Simple ECG functions account for this by requiring nearly overlapping terms in the linear combinations 
with large matching linear coefficients of opposite sign.
This near-linear dependency in the basis complicates optimization and yields 
numerically unstable eigensystems with ill-conditioned Hamiltonian matrices.
Conversely, we did not encounter such severe near-linear dependencies with apzFECG functions 
because these functions can effectively separate the proton densities along an axis.

\section{Numerical results} \label{SEC:numericalresults}

The formulae derived we implemented in a \texttt{C++} computer program. 
These analytical expressions allow us to calculate matrix elements reliably.
Other sources of error such as numerical integration or truncation of infinite series 
are eliminated by our approach. 

As test examples for the novel basis function presented in this work we chose the 
dihydrogen molecular ion, H$_2^+=\{$p$^+,$p$^+,$e$^-\}$, and dihydrogen, H$_2=\{$p$^+,$p$^+,$e$^-,$e$^-\}$
treated explicitly as three and four-particle systems, respectively.
The Born-Oppenheimer approximation is not invoked, i.e., nuclei and electrons are described on equal footing.
The energies obtained for the first three rotational states are shown in 
Tables \ref{TAB:proj_zFECG_1} and \ref{TAB:proj_zFECG_3}, respectively. 
For each state, we optimized a different basis sets consisting of $400$ and $600$ zFECG 
functions, respectively. Matrix elements were calculated as discussed in Sec.~\ref{SEC:matrixelements} 
where the projection operator was applied to the ket function.
The virial coefficient, $\eta=|1+\langle\Psi|\hat{V}|\Psi\rangle/(2\langle\Psi|\hat{T}|\Psi\rangle)|$
vanishes for the exact solution \cite{suzukivarga}, so that it represents a diagnostic for the overall 
quality of the variationally optimized wave function.
The basis set size was gradually increased following the competitive selection method \cite{suzukivarga}
for which the newer basis functions entering the basis set are selected from a large pool of
randomly generated trial functions.
A simultaneous refinement of the non-linear variational parameters was crucial to achieve efficient energy convergence. 
This optimization problem of minimizing the energy with respect to the set of non-linear parameters 
is a difficult problem as the objective function is non-convex, non-separable, and often (Sec.~\ref{SEC:NumStab}) ill-conditioned.
We relied on two derivative-free algorithms: the Subplex algorithm by Rowan \cite{Subplex}
and the Principal Axis method discussed by Brent \cite{Praxis}.
In our computer implementation of both methods, we used the \texttt{NLopt} package \cite{nlopt}.
We employed our multi-channel optimization approach presented in our previous work \cite{Muolo2018a}
and we have included every possible set of Jacobi coordinates, the heavy-particle-centered coordinates,
and the center-of-mass-centered coordinates. 
The construction of the Gaussian parameters through different $U_a^{\text{TICC}}$ maps
allows us to explore the parameter space faster and to describe different groupings of the particles 
with the most appropriate TICC set.
These calculations were carried out using
message passing interface (MPI) parallelization on six multiprocessor computer platforms
(AMD Opteron$^{\texttt{TM}}$ Processor 6376). 

We compare the results for H$_2^+$ and H$_2$ with Ref.~\cite{H2+_1} and \cite{Pachucki2009},
respectively.
Earlier results obtained with unprojected FECG and numerically projected FECG functions
(with three-dimensional shifted centers) for H$_2$ with a basis set size of $N_b=1560$
are $1.162739$ E$_{\text{h}}$ and $1.163998$ E$_{\text{h}}$, respectively \cite{Muolo2018b}.
The wall time of these earlier calculations was about three months.
Our best result with only $600$ linearly combined apzFECGs for the rotational ground state 
of H$_2$ is $-1.16402502482$ E$_{\text{h}}$.
Accordingly, the wall time of the calculation was reduced to about two months yielding a result of higher accuracy.
Investigating the results in Tables \ref{TAB:proj_zFECG_1} and \ref{TAB:proj_zFECG_3}, 
we observe that the energies are well converged with the number of basis functions. 
The optimized basis-function parameters are deposited in the supplementary material. 

\begin{table*}[h]
\centering
\caption{\label{TAB:proj_zFECG_1}
         \footnotesize{ Nonrelativistic energies of H$_2^+=\{$p$^+,$p$^+,$e$^-\}$, compared with results from Ref.~\cite{H2+_1} in the last column. The calculations include all possible Jacobi coordinates, 
the heavy-particle-centered, and the center-of-mass-centered coordinate sets.}
        }
\footnotesize{
  \begin{tabular} { @{\hspace{1.0mm}} c @{\hspace{4.0mm}} c @{\hspace{4.0mm}} c @{\hspace{4.0mm}} c @{\hspace{4.0mm}} c @{\hspace{1.0mm}} }
  \hline
  \hline
  & $\langle\hat{H}\rangle/{\text E}_{\text h}$ $(N_b=400)$ & 
    $\eta$ &
    $\langle\hat{H}\rangle_{\text{Ref.}}/{\text E}_{\text h}$ $(N_b=4000)$ & 
    $^a\Delta E/n{\text E}_{\text h}$ \\
  \hline
  $N=0$  &  $-0.597139062111$  &  $10^{-9}$  &  $-0.597139063079$  &  $-0.968$   \\
  $N=1$  &  $-0.596873736772$  &  $10^{-9}$  &  $-0.596873738784$  &  $-2.012$   \\
  $N=2$  &  $-0.596345204133$  &  $10^{-9}$  &  $-0.596345205489$  &  $-1.356$   \\
  \hline
  \end{tabular}
}
\caption*{ \footnotesize{ $^a\Delta E=E({\text{Ref.}}$~\cite{H2+_1}$)-\langle\hat{H}\rangle$  
} }
\end{table*}

\begin{table*}[h]
\centering
\caption{\label{TAB:proj_zFECG_3}
         \footnotesize{ Nonrelativistic energies of H$_2=\{$p$^+,$p$^+,$e$^-,$e$^-\}$, compared with results from Ref.~\cite{Pachucki2009} in the last column. The calculations include all possible Jacobi coordinates, 
the heavy-particle-centered, and the center-of-mass-centered coordinate sets. }
        }
\footnotesize{
  \begin{tabular} { @{\hspace{1.0mm}} c @{\hspace{4.0mm}} c @{\hspace{4.0mm}} c @{\hspace{4.0mm}} c @{\hspace{4.0mm}} c @{\hspace{1.0mm}} }
  \hline
  \hline
  & $\langle\hat{H}\rangle/{\text E}_{\text h}$ $(N_b=600)$ & 
    $\eta$ &
    $\langle\hat{H}\rangle_{\text{Ref.}}/{\text E}_{\text h}$ $(N_b=4200)$ & 
    $^a\Delta E/n{\text E}_{\text h}$ \\
  \hline
  $N=0$  &  $-1.16402502482$  &  $10^{-8}$  &  $-1.164025031$  &  $-6.18$   \\
  $N=1$  &  $-1.16348516709$  &  $10^{-8}$  &  $-1.163485173$  &  $-5.91$   \\
  $N=2$  &  $-1.16241040566$  &  $10^{-7}$  &  $-1.162410409$  &  $-3.34$   \\
  \hline
  \end{tabular}
}
\caption*{ \footnotesize{ $^a\Delta E=E({\text{Ref.}}$~\cite{Pachucki2009}$)-\langle\hat{H}\rangle$  
} }
\end{table*}

\section{Conclusions}

Projection techniques increase the effectiveness of variational 
basis function optimization carried out in the desired eigenspace. 
The formalism developed in this paper analytically solves the projection based approach 
for the subset of explicitly correlated floating Gaussian functions
having shift vectors aligned on one axis. 
We have derived analytical expressions of important matrix elements for 
projected zFECGs with arbitrary angular momentum and parity configurations. 
The resulting analytically projected zFECGs can potentially target any rotational state.
This can be done efficiently because they are eigenfunctions of 
the total (nuclei plus electrons) squared spatial angular momentum operator $\hat{N}^2$
with eigenvalue $N$ and of $\hat{N}_z$ with eigenvalue $M_N=0$.
Since only states with zero total spatial angular momentum projection onto the $z$ axis
can be accessed, among the $2N+1$ degenerate states with $M_N=-N,\ldots,+N$, 
these functions are not suited in applications for which these degeneracies
are lifted, e.g., in the presence of external magnetic fields.
Despite this limitation, projected zFECGs address the problem of targeting rotationally 
excited states exactly, whereas other explicitly correlated basis functions either 
specialize on one specific $N$ considering only lowest-order angular momentum couplings
for the ease of the Hamiltonian matrix elements,
or resemble the correct partial wave decomposition only for very high linear combinations 
and in the variational limit with the so-called global vector representation.
The numerical examples presented demonstrate the correctness of the derived formulae 
and the applicability of the approach to excited rotational states of small molecules. 

Particularly interesting will be the application
of our new analytical projection method to shift vectors lying on a plane 
and the extension to floating Gaussian functions with pre-exponential factors 
which can well represent the radial nodes of, for example, pure vibrational states.
Such calculations are beyond the scope of the present paper and are therefore deferred
to future work.

\section*{Acknowledgments}

This work has been financially supported by ETH Zurich.

\appendix

\section{List of integrals}

This appendix reviews the solutions to the principal integrals of the overlap, kinetic, 
and total angular momentum squared integral matrix elements for apzFECG functions.
All formulas have been checked for consistency against multiple implementations 
and known special cases ($\bm{s}_I=0$, $C=0$).
The list of analytical solutions to the principal integrals for $N\in [0,10]$ is as follows:

\small
\begin{flalign}
\Upsilon^{0}_{0} =& ~
\displaystyle{\frac{2}{\pi C}} \sinh (C) 
&&
\end{flalign}
\begin{flalign}
\Upsilon^{1}_{0} =& ~
\displaystyle{\frac{2}{\pi C}} \cosh (C) - \displaystyle{\frac{2}{\pi C^2}} \sinh (C) 
&&
\end{flalign}
\begin{flalign}
\Upsilon^{2}_{0} =& ~
\displaystyle{\frac{2}{\pi C^3}} \Big[ (C^2+3) \sinh (C)-3 C \cosh (C)\Big]
&&
\end{flalign}
\begin{flalign}
\Upsilon^{3}_{0} =& ~
\displaystyle{\frac{2}{\pi C^4}}
\Big[C (C^2+15) \cosh (C)-3 (2 C^2+5) \sinh (C)\Big]
&&
\end{flalign}
\begin{flalign}
\Upsilon^{4}_{0} =& ~
\displaystyle{\frac{2}{\pi C^5}}
\Big[ (C^4+45 C^2+105) \sinh (C)-5 C (2 C^2+21) \cosh (C)\Big]
&&
\end{flalign}
\begin{flalign}
\Upsilon^{5}_{0} =& ~
\displaystyle{\frac{2}{\pi C^6}}
\Big[C (C^4+105 C^2+945) \cosh (C) -15 (C^4+28 C^2+63) \sinh (C)\Big]
&&
\end{flalign}
\begin{flalign}
\Upsilon^{6}_{0} =& ~
\displaystyle{\frac{2}{\pi C^7}}
\Big[ (C^6+210 C^4+4725 C^2+10395) \sinh (C) -21 C (C^4+60 C^2+495) \cosh (C)\Big]
&&
\end{flalign}
\begin{flalign}
\Upsilon^{7}_{0} =& ~
\displaystyle{\frac{2}{\pi C^8}}
\Big[C (C^6+378 C^4+17325 C^2+135135) \cosh (C)
\nonumber \\[0.2cm] 
& -7 (4 C^6+450 C^4+8910 C^2+19305) \sinh (C)\Big]
&&
\end{flalign}
\begin{flalign}
\Upsilon^{8}_{0} =& ~
\displaystyle{\frac{2}{\pi C^9}}
\Big[ (C^8+630 C^6+51975 C^4+945945 C^2+2027025) \sinh (C)
\nonumber \\[0.2cm]
& -9 C (4 C^6+770 C^4+30030 C^2+225225) \cosh (C)\Big]
&&
\end{flalign}
\begin{flalign}
\Upsilon^{9}_{0} =& ~
\displaystyle{\frac{2}{\pi C^{10}}}
\Big[C (C^8+990 C^6+135135 C^4+4729725 C^2+34459425) \cosh (C)
\nonumber \\[0.2cm] 
& -45 (C^8+308 C^6+21021 C^4+360360 C^2+765765) \sinh (C)\Big]
&&
\end{flalign}
\begin{flalign}
\Upsilon^{10}_{0} =& ~
\displaystyle{\frac{2}{\pi C^{11}}}
\Big[ (C^{10}+1485 C^8+315315 C^6+18918900 C^4+310134825 C^2+654729075) \sinh (C)
\nonumber \\[0.2cm]
& -55 C (C^8+468 C^6+51597 C^4+1670760 C^2+11904165) \cosh (C)\Big]
&&
\end{flalign}

\vspace{0.2cm}

\begin{flalign}
\Sigma^{0}_{0} =& ~
\displaystyle{\frac{2}{\pi C^2}} \Big[\sinh(C) (C\omega -\sigma)+C\sigma\cosh(C)\Big]  
&&
\end{flalign}
\begin{flalign}
\Sigma^{1}_{0} =& ~
\displaystyle{\frac{2}{\pi C^3}} \Big[\sinh(C) \left(\left(C^2+2\right) \sigma -C\omega \right)+C\cosh(C) (C\omega -2\sigma)\Big]  
&&
\end{flalign}
\begin{flalign}
\Sigma^{2}_{0} =& ~
\displaystyle{\frac{2}{\pi C^4}} \Big[\sinh(C) \left(C \left(C^2+3\right) \omega -\left(4 C^2+9\right) \sigma\right) 
&&
\end{flalign}
\begin{flalign}
\Sigma^{3}_{0} =& ~
\displaystyle{\frac{2}{\pi  C^5}}
\Big[C \cosh (C) \Big(C \left(C^2+15\right) \omega -\left(7 C^2+60\right) \sigma \Big) 
\nonumber \\[0.2cm]
& +\sinh (C) \Big(\left(C^4+27 C^2+60\right) \sigma -3 C \left(2 C^2+5\right) \omega \Big) \Big]
&&
\end{flalign}
\begin{flalign}
\Sigma^{4}_{0} =& ~
\displaystyle{\frac{2}{\pi  C^6}}
\Big[\sinh (C) \Big( C \left(C^4+45 C^2+105\right) \omega -\left(11 C^4+240
C^2+525\right) \sigma \Big) 
\nonumber \\[0.2cm]
& +C \cosh (C) \Big(\left(C^4+65 C^2+525\right) \sigma -5 C \left(2 C^2+21\right) \omega \Big) \Big]
&&
\end{flalign}
\begin{flalign}
\Sigma^{5}_{0} =& ~
\displaystyle{\frac{2}{\pi  C^7}}
\Big[ C \cosh (C) \Big( C \left(C^4+105 C^2+945\right) \omega -\left(16 C^4+735
C^2+5670\right) \sigma \Big) 
\nonumber \\[0.2cm] 
& +\sinh (C) \Big(\left(C^6+135 C^4+2625 C^2+5670\right) \sigma 
-15 C \left(C^4+28 C^2+63\right) \omega \Big) \Big]
&&
\end{flalign}
\begin{flalign}
\Sigma^{6}_{0} =& ~
\displaystyle{\frac{2}{\pi  C^8}}
\Big( \sinh (C) \Big( C \left(C^6+210 C^4+4725 C^2+10395\right) \omega 
\nonumber \\[0.2cm] 
& -\left(22 C^6+1890 C^4+34020 C^2+72765\right) \sigma \Big) 
\nonumber \\[0.2cm] 
& +C \cosh (C) \Big(\left(C^6+252 C^4+9765 C^2+72765\right) \sigma 
-21 C \left(C^4+60 C^2+495\right) \omega \Big)\Big]
&&
\end{flalign}
\begin{flalign}
\Sigma^{7}_{0} =& ~
\displaystyle{\frac{2}{\pi  C^9}}
\Big[C \cosh (C) \Big(C \left(C^6+378 C^4+17325 C^2+135135\right) \omega 
\nonumber \\[0.2cm]
& -\left(29 C^6+4284 C^4+148995 C^2+1081080\right) \sigma \Big)
\nonumber \\[0.2cm]
& +\sinh (C) \Big(\left(C^8+434 C^6+29925 C^4+509355 C^2+1081080\right) \sigma 
\nonumber \\[0.2cm]
& -7 C \left(4 C^6+450 C^4+8910 C^2+19305\right) \omega \Big) \Big]
&&
\end{flalign}
\begin{flalign}
\Sigma^{8}_{0} =& ~
\displaystyle{\frac{2}{\pi  C^{10}}}
\Big( \sinh (C) \Big(C \left(C^8+630 C^6+51975 C^4+945945 C^2+2027025\right) \omega
\nonumber \\[0.2cm]
& -\left(37 C^8+8820 C^6+530145 C^4+8648640 C^2+18243225\right) \sigma \Big)
\nonumber \\[0.2cm]
& +C \cosh (C) \Big(\left(C^8+702 C^6+79695 C^4+2567565 C^2+18243225\right) \sigma 
\nonumber \\[0.2cm]
& -9 C \left(4 C^6+770 C^4+30030 C^2+225225\right) \omega \Big) \Big]
&&
\end{flalign}
\begin{flalign}
\Sigma^{9}_{0} =& ~
\displaystyle{\frac{2}{\pi  C^{11}}}
\Big[C \cosh (C) \Big(C \left(C^8+990 C^6+135135 C^4+4729725 C^2+34459425\right) \omega 
\nonumber \\[0.2cm] 
& -\left(46 C^8+16830 C^6+1621620 C^4+49324275 C^2+344594250\right) \sigma \Big)
\nonumber \\[0.2cm] 
& +\sinh (C) \Big( \left(C^{10}+1080 C^8+190575 C^6+10405395 C^4+164189025
C^2+344594250\right) \sigma 
\nonumber \\[0.2cm] 
& -45 C \left(C^8+308 C^6+21021 C^4+360360 C^2+765765\right) \omega 
\Big) \Big]
&&
\end{flalign}
\begin{flalign}
\Sigma^{10}_{0} =& ~
\displaystyle{\frac{2}{\pi  C^{12}}}
\Big(\sinh (C) \Big( C (C^{10}+1485 C^8+315315 C^6+18918900 C^4
\nonumber \\[0.2cm] 
& +310134825 C^2+654729075) \omega -(56 C^{10}+30195 C^8+4414410 C^6+224324100 C^4
\nonumber \\[0.2cm]
& +3445942500 C^2+7202019825) \sigma \Big)
+C \cosh (C) \Big( (C^{10}+1595 C^8+418275 C^6
\nonumber \\[0.2cm]
& +35945910 C^4+1045269225 C^2+7202019825) \sigma 
\nonumber \\[0.2cm]
& -55 C \left(C^8+468 C^6+51597 C^4+1670760 C^2+11904165\right) \omega \Big) \Big)
&&
\end{flalign}
\begin{flalign}
\Xi^{0}_{0} =& ~ 0                
&&
\end{flalign}
\begin{flalign}
\Xi^{1}_{0} =& ~ 2 \,\, \Upsilon^1_0                
&&
\end{flalign}
\begin{flalign}
\Xi^{2}_{0} =& ~ 6 \,\, \Upsilon^2_0               
&&
\end{flalign}
\begin{flalign}
\Xi^{3}_{0} =& ~ 12 \,\, \Upsilon^3_0         
&&
\end{flalign}
\begin{flalign}
\Xi^{4}_{0} =& ~ 20 \,\, \Upsilon^4_0         
&&
\end{flalign}
\begin{flalign}
\Xi^{5}_{0} =& ~ 30 \,\, \Upsilon^5_0         
&&
\end{flalign}

\normalsize


\begin{thebibliography}{54}%
\makeatletter
\providecommand \@ifxundefined [1]{%
 \@ifx{#1\undefined}
}%
\providecommand \@ifnum [1]{%
 \ifnum #1\expandafter \@firstoftwo
 \else \expandafter \@secondoftwo
 \fi
}%
\providecommand \@ifx [1]{%
 \ifx #1\expandafter \@firstoftwo
 \else \expandafter \@secondoftwo
 \fi
}%
\providecommand \natexlab [1]{#1}%
\providecommand \enquote  [1]{``#1''}%
\providecommand \bibnamefont  [1]{#1}%
\providecommand \bibfnamefont [1]{#1}%
\providecommand \citenamefont [1]{#1}%
\providecommand \href@noop [0]{\@secondoftwo}%
\providecommand \href [0]{\begingroup \@sanitize@url \@href}%
\providecommand \@href[1]{\@@startlink{#1}\@@href}%
\providecommand \@@href[1]{\endgroup#1\@@endlink}%
\providecommand \@sanitize@url [0]{\catcode `\\12\catcode `\$12\catcode
  `\&12\catcode `\#12\catcode `\^12\catcode `\_12\catcode `\%12\relax}%
\providecommand \@@startlink[1]{}%
\providecommand \@@endlink[0]{}%
\providecommand \url  [0]{\begingroup\@sanitize@url \@url }%
\providecommand \@url [1]{\endgroup\@href {#1}{\urlprefix }}%
\providecommand \urlprefix  [0]{URL }%
\providecommand \Eprint [0]{\href }%
\providecommand \doibase [0]{http://dx.doi.org/}%
\providecommand \selectlanguage [0]{\@gobble}%
\providecommand \bibinfo  [0]{\@secondoftwo}%
\providecommand \bibfield  [0]{\@secondoftwo}%
\providecommand \translation [1]{[#1]}%
\providecommand \BibitemOpen [0]{}%
\providecommand \bibitemStop [0]{}%
\providecommand \bibitemNoStop [0]{.\EOS\space}%
\providecommand \EOS [0]{\spacefactor3000\relax}%
\providecommand \BibitemShut  [1]{\csname bibitem#1\endcsname}%
\let\auto@bib@innerbib\@empty
\bibitem [{\citenamefont {Boys}(1960)}]{ECG-history1960}%
  \BibitemOpen
  \bibfield  {author} {\bibinfo {author} {\bibfnamefont {S.~F.}\ \bibnamefont
  {Boys}},\ }\href {\doibase 10.1098/rspa.1960.0195} {\bibfield  {journal}
  {\bibinfo  {journal} {Proc. R. Soc. London, Ser. A}\ }\textbf {\bibinfo
  {volume} {258}},\ \bibinfo {pages} {402} (\bibinfo {year}
  {1960})}\BibitemShut {NoStop}%
\bibitem [{\citenamefont {Singer}(1960)}]{ECG-history1960_1}%
  \BibitemOpen
  \bibfield  {author} {\bibinfo {author} {\bibfnamefont {K.}~\bibnamefont
  {Singer}},\ }\href {\doibase 10.1098/rspa.1960.0196} {\bibfield  {journal}
  {\bibinfo  {journal} {Proc. R. Soc. London, Ser. A}\ }\textbf {\bibinfo
  {volume} {258}},\ \bibinfo {pages} {412} (\bibinfo {year}
  {1960})}\BibitemShut {NoStop}%
\bibitem [{\citenamefont {Kukulin}\ and\ \citenamefont
  {Krasnopol'sky}(1977)}]{svm-history1977}%
  \BibitemOpen
  \bibfield  {author} {\bibinfo {author} {\bibfnamefont {V.~I.}\ \bibnamefont
  {Kukulin}}\ and\ \bibinfo {author} {\bibfnamefont {V.~M.}\ \bibnamefont
  {Krasnopol'sky}},\ }\href@noop {} {\bibfield  {journal} {\bibinfo  {journal}
  {J. Phys. G}\ }\textbf {\bibinfo {volume} {3}},\ \bibinfo {pages} {795}
  (\bibinfo {year} {1977})}\BibitemShut {NoStop}%
\bibitem [{\citenamefont {Thakkar}\ and\ \citenamefont
  {Smith}(1977)}]{randTempe-1}%
  \BibitemOpen
  \bibfield  {author} {\bibinfo {author} {\bibfnamefont {A.~J.}\ \bibnamefont
  {Thakkar}}\ and\ \bibinfo {author} {\bibfnamefont {V.~H.}\ \bibnamefont
  {Smith}},\ }\href {\doibase 10.1103/PhysRevA.15.1} {\bibfield  {journal}
  {\bibinfo  {journal} {Phys. Rev. A}\ }\textbf {\bibinfo {volume} {15}},\
  \bibinfo {pages} {1} (\bibinfo {year} {1977})}\BibitemShut {NoStop}%
\bibitem [{\citenamefont {Poshusta}(1978)}]{FECG-integrals-advances1978}%
  \BibitemOpen
  \bibfield  {author} {\bibinfo {author} {\bibfnamefont {R.~D.}\ \bibnamefont
  {Poshusta}},\ }\href@noop {} {\bibfield  {journal} {\bibinfo  {journal} {Int.
  J. Quantum Chem.}\ }\textbf {\bibinfo {volume} {13}},\ \bibinfo {pages} {27}
  (\bibinfo {year} {1978})}\BibitemShut {NoStop}%
\bibitem [{\citenamefont {Jeziorski}\ and\ \citenamefont
  {Szalewicz}(1979)}]{Szalewicz_1979}%
  \BibitemOpen
  \bibfield  {author} {\bibinfo {author} {\bibfnamefont {B.}~\bibnamefont
  {Jeziorski}}\ and\ \bibinfo {author} {\bibfnamefont {K.}~\bibnamefont
  {Szalewicz}},\ }\href {\doibase 10.1103/PhysRevA.19.2360} {\bibfield
  {journal} {\bibinfo  {journal} {Phys. Rev. A}\ }\textbf {\bibinfo {volume}
  {19}},\ \bibinfo {pages} {2360} (\bibinfo {year} {1979})}\BibitemShut
  {NoStop}%
\bibitem [{\citenamefont {Szalewicz}\ \emph {et~al.}(1983)\citenamefont
  {Szalewicz}, \citenamefont {Jeziorski}, \citenamefont {Monkhorst},\ and\
  \citenamefont {Zabolitzky}}]{Szalewicz_II_1983}%
  \BibitemOpen
  \bibfield  {author} {\bibinfo {author} {\bibfnamefont {K.}~\bibnamefont
  {Szalewicz}}, \bibinfo {author} {\bibfnamefont {B.}~\bibnamefont
  {Jeziorski}}, \bibinfo {author} {\bibfnamefont {H.~J.}\ \bibnamefont
  {Monkhorst}}, \ and\ \bibinfo {author} {\bibfnamefont {J.~G.}\ \bibnamefont
  {Zabolitzky}},\ }\href {\doibase 10.1063/1.445672} {\bibfield  {journal}
  {\bibinfo  {journal} {J.\,Chem.\,Phys.}\ }\textbf {\bibinfo {volume} {79}},\
  \bibinfo {pages} {5543} (\bibinfo {year} {1983})}\BibitemShut {NoStop}%
\bibitem [{\citenamefont {Alexander}\ \emph {et~al.}(1986)\citenamefont
  {Alexander}, \citenamefont {Monkhorst},\ and\ \citenamefont
  {Szalewicz}}]{randTempe-2}%
  \BibitemOpen
  \bibfield  {author} {\bibinfo {author} {\bibfnamefont {S.~A.}\ \bibnamefont
  {Alexander}}, \bibinfo {author} {\bibfnamefont {H.~J.}\ \bibnamefont
  {Monkhorst}}, \ and\ \bibinfo {author} {\bibfnamefont {K.}~\bibnamefont
  {Szalewicz}},\ }\href {\doibase 10.1063/1.451543} {\bibfield  {journal}
  {\bibinfo  {journal} {J.\,Chem.\,Phys.}\ }\textbf {\bibinfo {volume} {85}},\
  \bibinfo {pages} {5821} (\bibinfo {year} {1986})}\BibitemShut {NoStop}%
\bibitem [{\citenamefont {Alexander}\ \emph {et~al.}(1987)\citenamefont
  {Alexander}, \citenamefont {Monkhorst},\ and\ \citenamefont
  {Szalewicz}}]{randTempe-3}%
  \BibitemOpen
  \bibfield  {author} {\bibinfo {author} {\bibfnamefont {S.~A.}\ \bibnamefont
  {Alexander}}, \bibinfo {author} {\bibfnamefont {H.~J.}\ \bibnamefont
  {Monkhorst}}, \ and\ \bibinfo {author} {\bibfnamefont {K.}~\bibnamefont
  {Szalewicz}},\ }\href {\doibase 10.1063/1.452951} {\bibfield  {journal}
  {\bibinfo  {journal} {J.\,Chem.\,Phys.}\ }\textbf {\bibinfo {volume} {87}},\
  \bibinfo {pages} {3976} (\bibinfo {year} {1987})}\BibitemShut {NoStop}%
\bibitem [{\citenamefont {Alexander}\ \emph {et~al.}(1990)\citenamefont
  {Alexander}, \citenamefont {Monkhorst}, \citenamefont {Roeland},\ and\
  \citenamefont {Szalewicz}}]{Szalewicz_1990}%
  \BibitemOpen
  \bibfield  {author} {\bibinfo {author} {\bibfnamefont {S.~A.}\ \bibnamefont
  {Alexander}}, \bibinfo {author} {\bibfnamefont {H.~J.}\ \bibnamefont
  {Monkhorst}}, \bibinfo {author} {\bibfnamefont {R.}~\bibnamefont {Roeland}},
  \ and\ \bibinfo {author} {\bibfnamefont {K.}~\bibnamefont {Szalewicz}},\
  }\href {\doibase 10.1063/1.458755} {\bibfield  {journal} {\bibinfo  {journal}
  {J.\,Chem.\,Phys.}\ }\textbf {\bibinfo {volume} {93}},\ \bibinfo {pages}
  {4230} (\bibinfo {year} {1990})}\BibitemShut {NoStop}%
\bibitem [{\citenamefont {Cencek}\ and\ \citenamefont
  {Rychlewski}(1993)}]{ECG-history1993}%
  \BibitemOpen
  \bibfield  {author} {\bibinfo {author} {\bibfnamefont {W.}~\bibnamefont
  {Cencek}}\ and\ \bibinfo {author} {\bibfnamefont {J.}~\bibnamefont
  {Rychlewski}},\ }\href {\doibase http://dx.doi.org/10.1063/1.464293}
  {\bibfield  {journal} {\bibinfo  {journal} {J.\,Chem.\,Phys.}\ }\textbf
  {\bibinfo {volume} {98}},\ \bibinfo {pages} {1252} (\bibinfo {year}
  {1993})}\BibitemShut {NoStop}%
\bibitem [{\citenamefont {Yan}\ and\ \citenamefont {Drake}(1997)}]{Drake1997}%
  \BibitemOpen
  \bibfield  {author} {\bibinfo {author} {\bibfnamefont {Z.-C.}\ \bibnamefont
  {Yan}}\ and\ \bibinfo {author} {\bibfnamefont {G.~W.~F.}\ \bibnamefont
  {Drake}},\ }\href@noop {} {\bibfield  {journal} {\bibinfo  {journal} {J.
  Phys. B}\ }\textbf {\bibinfo {volume} {30}},\ \bibinfo {pages} {4723}
  (\bibinfo {year} {1997})}\BibitemShut {NoStop}%
\bibitem [{\citenamefont {Korobov}(2000)}]{Korobov2000}%
  \BibitemOpen
  \bibfield  {author} {\bibinfo {author} {\bibfnamefont {V.~I.}\ \bibnamefont
  {Korobov}},\ }\href {\doibase 10.1103/PhysRevA.61.064503} {\bibfield
  {journal} {\bibinfo  {journal} {Phys. Rev. A}\ }\textbf {\bibinfo {volume}
  {61}},\ \bibinfo {pages} {064503} (\bibinfo {year} {2000})}\BibitemShut
  {NoStop}%
\bibitem [{\citenamefont {Cafiero}\ \emph {et~al.}(2003)\citenamefont
  {Cafiero}, \citenamefont {Bubin},\ and\ \citenamefont
  {Adamowicz}}]{Adamowicz2003a}%
  \BibitemOpen
  \bibfield  {author} {\bibinfo {author} {\bibfnamefont {M.}~\bibnamefont
  {Cafiero}}, \bibinfo {author} {\bibfnamefont {S.}~\bibnamefont {Bubin}}, \
  and\ \bibinfo {author} {\bibfnamefont {L.}~\bibnamefont {Adamowicz}},\
  }\href@noop {} {\bibfield  {journal} {\bibinfo  {journal} {Phys. Chem. Chem.
  Phys.}\ }\textbf {\bibinfo {volume} {5}},\ \bibinfo {pages} {1491} (\bibinfo
  {year} {2003})}\BibitemShut {NoStop}%
\bibitem [{\citenamefont {M\'atyus}\ and\ \citenamefont
  {Reiher}(2012)}]{Matyus2012}%
  \BibitemOpen
  \bibfield  {author} {\bibinfo {author} {\bibfnamefont {E.}~\bibnamefont
  {M\'atyus}}\ and\ \bibinfo {author} {\bibfnamefont {M.}~\bibnamefont
  {Reiher}},\ }\href@noop {} {\bibfield  {journal} {\bibinfo  {journal}
  {J.\,Chem.\,Phys.}\ }\textbf {\bibinfo {volume} {137}},\ \bibinfo {pages} {024104} (\bibinfo {year}
  {2012})}\BibitemShut {NoStop}%
\bibitem [{\citenamefont {Pachucki}(2012)}]{Pachucki2012}%
  \BibitemOpen
  \bibfield  {author} {\bibinfo {author} {\bibfnamefont {K.}~\bibnamefont
  {Pachucki}},\ }\href {\doibase 10.1103/PhysRevA.86.052514} {\bibfield
  {journal} {\bibinfo  {journal} {Phys. Rev. A}\ }\textbf {\bibinfo {volume}
  {86}},\ \bibinfo {pages} {052514} (\bibinfo {year} {2012})}\BibitemShut
  {NoStop}%
\bibitem [{\citenamefont {Bubin}\ \emph {et~al.}(2013)\citenamefont {Bubin},
  \citenamefont {Pavanello}, \citenamefont {Tung}, \citenamefont {Sharkey},\
  and\ \citenamefont {Adamowicz}}]{Adamowicz2013}%
  \BibitemOpen
  \bibfield  {author} {\bibinfo {author} {\bibfnamefont {S.}~\bibnamefont
  {Bubin}}, \bibinfo {author} {\bibfnamefont {M.}~\bibnamefont {Pavanello}},
  \bibinfo {author} {\bibfnamefont {W.-C.}\ \bibnamefont {Tung}}, \bibinfo
  {author} {\bibfnamefont {K.~L.}\ \bibnamefont {Sharkey}}, \ and\ \bibinfo
  {author} {\bibfnamefont {L.}~\bibnamefont {Adamowicz}},\ }\href {\doibase
  10.1021/cr200419d} {\bibfield  {journal} {\bibinfo  {journal} {Chem. Rev.}\
  }\textbf {\bibinfo {volume} {113}},\ \bibinfo {pages} {36} (\bibinfo {year}
  {2013})}\BibitemShut {NoStop}%
\bibitem [{\citenamefont {Puchalski}\ \emph {et~al.}(2015)\citenamefont
  {Puchalski}, \citenamefont {Komasa},\ and\ \citenamefont
  {Pachucki}}]{Pachucki2015}%
  \BibitemOpen
  \bibfield  {author} {\bibinfo {author} {\bibfnamefont {M.}~\bibnamefont
  {Puchalski}}, \bibinfo {author} {\bibfnamefont {J.}~\bibnamefont {Komasa}}, \
  and\ \bibinfo {author} {\bibfnamefont {K.}~\bibnamefont {Pachucki}},\ }\href
  {\doibase 10.1103/PhysRevA.92.062501} {\bibfield  {journal} {\bibinfo
  {journal} {Phys. Rev. A}\ }\textbf {\bibinfo {volume} {92}},\ \bibinfo
  {pages} {062501} (\bibinfo {year} {2015})}\BibitemShut {NoStop}%
\bibitem [{\citenamefont {Hylleraas}(1928)}]{Hylleraas1928}%
  \BibitemOpen
  \bibfield  {author} {\bibinfo {author} {\bibfnamefont {E.~A.}\ \bibnamefont
  {Hylleraas}},\ }\href {\doibase 10.1007/BF01340013} {\bibfield  {journal}
  {\bibinfo  {journal} {Zeitschrift f{\"u}r Physik}\ }\textbf {\bibinfo
  {volume} {48}},\ \bibinfo {pages} {469} (\bibinfo {year} {1928})}\BibitemShut
  {NoStop}%
\bibitem [{\citenamefont {Hylleraas}(1929{\natexlab{a}})}]{Hylleraas1929a}%
  \BibitemOpen
  \bibfield  {author} {\bibinfo {author} {\bibfnamefont {E.~A.}\ \bibnamefont
  {Hylleraas}},\ }\href {\doibase 10.1007/BF01506430} {\bibfield  {journal}
  {\bibinfo  {journal} {Naturwissenschaften}\ }\textbf {\bibinfo {volume}
  {17}},\ \bibinfo {pages} {982} (\bibinfo {year}
  {1929}{\natexlab{a}})}\BibitemShut {NoStop}%
\bibitem [{\citenamefont {Hylleraas}(1929{\natexlab{b}})}]{Hylleraas1929b}%
  \BibitemOpen
  \bibfield  {author} {\bibinfo {author} {\bibfnamefont {E.~A.}\ \bibnamefont
  {Hylleraas}},\ }\href {\doibase 10.1007/BF01375457} {\bibfield  {journal}
  {\bibinfo  {journal} {Zeitschrift f{\"u}r Physik}\ }\textbf {\bibinfo
  {volume} {54}},\ \bibinfo {pages} {347} (\bibinfo {year}
  {1929}{\natexlab{b}})}\BibitemShut {NoStop}%
\bibitem [{\citenamefont {Hylleraas}(1930)}]{Hylleraas1930a}%
  \BibitemOpen
  \bibfield  {author} {\bibinfo {author} {\bibfnamefont {E.~A.}\ \bibnamefont
  {Hylleraas}},\ }\href {\doibase 10.1007/BF01421744} {\bibfield  {journal}
  {\bibinfo  {journal} {Zeitschrift f{\"u}r Physik}\ }\textbf {\bibinfo
  {volume} {63}},\ \bibinfo {pages} {291} (\bibinfo {year} {1930})}\BibitemShut
  {NoStop}%
\bibitem [{\citenamefont {Hylleraas}\ and\ \citenamefont
  {Undheim}(1930)}]{Hylleraas1930b}%
  \BibitemOpen
  \bibfield  {author} {\bibinfo {author} {\bibfnamefont {E.~A.}\ \bibnamefont
  {Hylleraas}}\ and\ \bibinfo {author} {\bibfnamefont {B.}~\bibnamefont
  {Undheim}},\ }\href {\doibase 10.1007/BF01397263} {\bibfield  {journal}
  {\bibinfo  {journal} {Zeitschrift f{\"u}r Physik}\ }\textbf {\bibinfo
  {volume} {65}},\ \bibinfo {pages} {759} (\bibinfo {year} {1930})}\BibitemShut
  {NoStop}%
\bibitem [{\citenamefont {Perkins}(1973)}]{Perkins1973}%
  \BibitemOpen
  \bibfield  {author} {\bibinfo {author} {\bibfnamefont {J.~F.}\ \bibnamefont
  {Perkins}},\ }\href {\doibase 10.1103/PhysRevA.8.700} {\bibfield  {journal}
  {\bibinfo  {journal} {Phys. Rev. A}\ }\textbf {\bibinfo {volume} {8}},\
  \bibinfo {pages} {700} (\bibinfo {year} {1973})}\BibitemShut {NoStop}%
\bibitem [{\citenamefont {Fromm}\ and\ \citenamefont
  {Hill}(1987)}]{Hyl3ele_1987}%
  \BibitemOpen
  \bibfield  {author} {\bibinfo {author} {\bibfnamefont {D.~M.}\ \bibnamefont
  {Fromm}}\ and\ \bibinfo {author} {\bibfnamefont {R.~N.}\ \bibnamefont
  {Hill}},\ }\href {\doibase 10.1103/PhysRevA.36.1013} {\bibfield  {journal}
  {\bibinfo  {journal} {Phys. Rev. A}\ }\textbf {\bibinfo {volume} {36}},\
  \bibinfo {pages} {1013} (\bibinfo {year} {1987})}\BibitemShut {NoStop}%
\bibitem [{\citenamefont {King}\ \emph {et~al.}(2011)\citenamefont {King},
  \citenamefont {Quicker},\ and\ \citenamefont {Langer}}]{Langer2011}%
  \BibitemOpen
  \bibfield  {author} {\bibinfo {author} {\bibfnamefont {F.~W.}\ \bibnamefont
  {King}}, \bibinfo {author} {\bibfnamefont {D.}~\bibnamefont {Quicker}}, \
  and\ \bibinfo {author} {\bibfnamefont {J.}~\bibnamefont {Langer}},\ }\href
  {\doibase 10.1063/1.3569565} {\bibfield  {journal} {\bibinfo  {journal}
  {J.\,Chem.\,Phys.}\ }\textbf {\bibinfo {volume} {134}},\ \bibinfo {pages}
  {124114} (\bibinfo {year} {2011})}\BibitemShut {NoStop}%
\bibitem [{\citenamefont {Mitroy}\ \emph {et~al.}(2013)\citenamefont {Mitroy},
  \citenamefont {Bubin}, \citenamefont {Horiuchi}, \citenamefont {Suzuki},
  \citenamefont {Adamowicz}, \citenamefont {Cencek}, \citenamefont {Szalewicz},
  \citenamefont {Komasa}, \citenamefont {Blume},\ and\ \citenamefont
  {Varga}}]{Adamowicz2013_rev}%
  \BibitemOpen
  \bibfield  {author} {\bibinfo {author} {\bibfnamefont {J.}~\bibnamefont
  {Mitroy}}, \bibinfo {author} {\bibfnamefont {S.}~\bibnamefont {Bubin}},
  \bibinfo {author} {\bibfnamefont {W.}~\bibnamefont {Horiuchi}}, \bibinfo
  {author} {\bibfnamefont {Y.}~\bibnamefont {Suzuki}}, \bibinfo {author}
  {\bibfnamefont {L.}~\bibnamefont {Adamowicz}}, \bibinfo {author}
  {\bibfnamefont {W.}~\bibnamefont {Cencek}}, \bibinfo {author} {\bibfnamefont
  {K.}~\bibnamefont {Szalewicz}}, \bibinfo {author} {\bibfnamefont
  {J.}~\bibnamefont {Komasa}}, \bibinfo {author} {\bibfnamefont
  {D.}~\bibnamefont {Blume}}, \ and\ \bibinfo {author} {\bibfnamefont
  {K.}~\bibnamefont {Varga}},\ }\href {\doibase 10.1103/RevModPhys.85.693}
  {\bibfield  {journal} {\bibinfo  {journal} {Rev. Mod. Phys.}\ }\textbf
  {\bibinfo {volume} {85}},\ \bibinfo {pages} {693} (\bibinfo {year}
  {2013})}\BibitemShut {NoStop}%
\bibitem [{\citenamefont {Varga}\ and\ \citenamefont
  {Suzuki}(1995)}]{Varga1995}%
  \BibitemOpen
  \bibfield  {author} {\bibinfo {author} {\bibfnamefont {K.}~\bibnamefont
  {Varga}}\ and\ \bibinfo {author} {\bibfnamefont {Y.}~\bibnamefont {Suzuki}},\
  }\href {\doibase 10.1103/PhysRevC.52.2885} {\bibfield  {journal} {\bibinfo
  {journal} {Phys. Rev. C}\ }\textbf {\bibinfo {volume} {52}},\ \bibinfo
  {pages} {2885} (\bibinfo {year} {1995})}\BibitemShut {NoStop}%
\bibitem [{\citenamefont {Varga}\ and\ \citenamefont
  {Suzuki}(1996)}]{Varga1996}%
  \BibitemOpen
  \bibfield  {author} {\bibinfo {author} {\bibfnamefont {K.}~\bibnamefont
  {Varga}}\ and\ \bibinfo {author} {\bibfnamefont {Y.}~\bibnamefont {Suzuki}},\
  }\href {http://www.ncbi.nlm.nih.gov/pubmed/9913088} {\bibfield  {journal}
  {\bibinfo  {journal} {Phys. Rev. A}\ }\textbf {\bibinfo {volume} {53}},\
  \bibinfo {pages} {1907} (\bibinfo {year} {1996})}\BibitemShut {NoStop}%
\bibitem [{\citenamefont {Komasa}\ and\ \citenamefont
  {Rychlewski}(2001)}]{Komasa2001}%
  \BibitemOpen
  \bibfield  {author} {\bibinfo {author} {\bibfnamefont {J.}~\bibnamefont
  {Komasa}}\ and\ \bibinfo {author} {\bibfnamefont {J.}~\bibnamefont
  {Rychlewski}},\ }\href {\doibase
  https://doi.org/10.1016/S0009-2614(01)00563-2} {\bibfield  {journal}
  {\bibinfo  {journal} {Chem.\,Phys.\,Lett.}\ }\textbf {\bibinfo {volume}
  {342}},\ \bibinfo {pages} {185 } (\bibinfo {year} {2001})}\BibitemShut
  {NoStop}%
\bibitem [{\citenamefont {Bubin}\ and\ \citenamefont
  {Adamowicz}(2008)}]{Adamowicz2008_zCECG}%
  \BibitemOpen
  \bibfield  {author} {\bibinfo {author} {\bibfnamefont {S.}~\bibnamefont
  {Bubin}}\ and\ \bibinfo {author} {\bibfnamefont {L.}~\bibnamefont
  {Adamowicz}},\ }\href {\doibase 10.1063/1.2894866} {\bibfield  {journal}
  {\bibinfo  {journal} {J.\,Chem.\,Phys.}\ }\textbf {\bibinfo {volume} {128}},\
  \bibinfo {pages} {114107} (\bibinfo {year} {2008})}\BibitemShut {NoStop}%
\bibitem [{\citenamefont {Sharkey}\ \emph {et~al.}(2009)\citenamefont
  {Sharkey}, \citenamefont {Pavanello}, \citenamefont {Bubin},\ and\
  \citenamefont {Adamowicz}}]{Adamowicz2009_N1}%
  \BibitemOpen
  \bibfield  {author} {\bibinfo {author} {\bibfnamefont {K.~L.}\ \bibnamefont
  {Sharkey}}, \bibinfo {author} {\bibfnamefont {M.}~\bibnamefont {Pavanello}},
  \bibinfo {author} {\bibfnamefont {S.}~\bibnamefont {Bubin}}, \ and\ \bibinfo
  {author} {\bibfnamefont {L.}~\bibnamefont {Adamowicz}},\ }\href {\doibase
  10.1103/PhysRevA.80.062510} {\bibfield  {journal} {\bibinfo  {journal} {Phys.
  Rev. A}\ }\textbf {\bibinfo {volume} {80}},\ \bibinfo {pages} {062510}
  (\bibinfo {year} {2009})}\BibitemShut {NoStop}%
\bibitem [{\citenamefont {Sharkey}\ \emph {et~al.}(2013)\citenamefont
  {Sharkey}, \citenamefont {Kirnosov},\ and\ \citenamefont
  {Adamowicz}}]{Adamowicz2013_N1}%
  \BibitemOpen
  \bibfield  {author} {\bibinfo {author} {\bibfnamefont {K.~L.}\ \bibnamefont
  {Sharkey}}, \bibinfo {author} {\bibfnamefont {N.}~\bibnamefont {Kirnosov}}, \
  and\ \bibinfo {author} {\bibfnamefont {L.}~\bibnamefont {Adamowicz}},\ }\href
  {\doibase 10.1063/1.4826450} {\bibfield  {journal} {\bibinfo  {journal}
  {J.\,Chem.\,Phys.}\ }\textbf {\bibinfo {volume} {139}},\ \bibinfo {pages}
  {164119} (\bibinfo {year} {2013})}\BibitemShut {NoStop}%
\bibitem [{\citenamefont {Kirnosov}\ \emph {et~al.}(2015)\citenamefont
  {Kirnosov}, \citenamefont {Sharkey},\ and\ \citenamefont
  {Adamowicz}}]{Adamowicz2015_zECG}%
  \BibitemOpen
  \bibfield  {author} {\bibinfo {author} {\bibfnamefont {N.}~\bibnamefont
  {Kirnosov}}, \bibinfo {author} {\bibfnamefont {K.~L.}\ \bibnamefont
  {Sharkey}}, \ and\ \bibinfo {author} {\bibfnamefont {L.}~\bibnamefont
  {Adamowicz}},\ }\href@noop {} {\bibfield  {journal} {\bibinfo  {journal} {J.
  Phys. B}\ }\textbf {\bibinfo {volume} {48}},\ \bibinfo {pages} {195101}
  (\bibinfo {year} {2015})}\BibitemShut {NoStop}%
\bibitem [{\citenamefont {Sharkey}\ \emph {et~al.}(2010)\citenamefont
  {Sharkey}, \citenamefont {Bubin},\ and\ \citenamefont
  {Adamowicz}}]{Adamowicz2010_N2}%
  \BibitemOpen
  \bibfield  {author} {\bibinfo {author} {\bibfnamefont {K.~L.}\ \bibnamefont
  {Sharkey}}, \bibinfo {author} {\bibfnamefont {S.}~\bibnamefont {Bubin}}, \
  and\ \bibinfo {author} {\bibfnamefont {L.}~\bibnamefont {Adamowicz}},\ }\href
  {\doibase 10.1063/1.3419931} {\bibfield  {journal} {\bibinfo  {journal}
  {J.\,Chem.\,Phys.}\ }\textbf {\bibinfo {volume} {132}},\ \bibinfo {pages}
  {184106} (\bibinfo {year} {2010})}\BibitemShut {NoStop}%
\bibitem [{\citenamefont {Sharkey}\ \emph
  {et~al.}(2011{\natexlab{a}})\citenamefont {Sharkey}, \citenamefont {Bubin},\
  and\ \citenamefont {Adamowicz}}]{Adamowicz2011_N2}%
  \BibitemOpen
  \bibfield  {author} {\bibinfo {author} {\bibfnamefont {K.~L.}\ \bibnamefont
  {Sharkey}}, \bibinfo {author} {\bibfnamefont {S.}~\bibnamefont {Bubin}}, \
  and\ \bibinfo {author} {\bibfnamefont {L.}~\bibnamefont {Adamowicz}},\ }\href
  {\doibase 10.1063/1.3523348} {\bibfield  {journal} {\bibinfo  {journal}
  {J.\,Chem.\,Phys.}\ }\textbf {\bibinfo {volume} {134}},\ \bibinfo {pages}
  {044120} (\bibinfo {year} {2011}{\natexlab{a}})}\BibitemShut {NoStop}%
\bibitem [{\citenamefont {Sharkey}\ \emph
  {et~al.}(2011{\natexlab{b}})\citenamefont {Sharkey}, \citenamefont {Bubin},\
  and\ \citenamefont {Adamowicz}}]{Adamowicz2011_N2_Ryd1}%
  \BibitemOpen
  \bibfield  {author} {\bibinfo {author} {\bibfnamefont {K.~L.}\ \bibnamefont
  {Sharkey}}, \bibinfo {author} {\bibfnamefont {S.}~\bibnamefont {Bubin}}, \
  and\ \bibinfo {author} {\bibfnamefont {L.}~\bibnamefont {Adamowicz}},\ }\href
  {\doibase 10.1063/1.3591836} {\bibfield  {journal} {\bibinfo  {journal}
  {J.\,Chem.\,Phys.}\ }\textbf {\bibinfo {volume} {134}},\ \bibinfo {pages}
  {194114} (\bibinfo {year} {2011}{\natexlab{b}})}\BibitemShut {NoStop}%
\bibitem [{\citenamefont {Sharkey}\ \emph
  {et~al.}(2011{\natexlab{c}})\citenamefont {Sharkey}, \citenamefont {Bubin},\
  and\ \citenamefont {Adamowicz}}]{Adamowicz2011_N2_Ryd2}%
  \BibitemOpen
  \bibfield  {author} {\bibinfo {author} {\bibfnamefont {K.~L.}\ \bibnamefont
  {Sharkey}}, \bibinfo {author} {\bibfnamefont {S.}~\bibnamefont {Bubin}}, \
  and\ \bibinfo {author} {\bibfnamefont {L.}~\bibnamefont {Adamowicz}},\ }\href
  {\doibase 10.1103/PhysRevA.83.012506} {\bibfield  {journal} {\bibinfo
  {journal} {Phys. Rev. A}\ }\textbf {\bibinfo {volume} {83}},\ \bibinfo
  {pages} {012506} (\bibinfo {year} {2011}{\natexlab{c}})}\BibitemShut
  {NoStop}%
\bibitem [{\citenamefont {Varga}\ \emph {et~al.}(1998)\citenamefont {Varga},
  \citenamefont {Suzuki},\ and\ \citenamefont {Usukura}}]{suzukivarga1998}%
  \BibitemOpen
  \bibfield  {author} {\bibinfo {author} {\bibfnamefont {K.}~\bibnamefont
  {Varga}}, \bibinfo {author} {\bibfnamefont {Y.}~\bibnamefont {Suzuki}}, \
  and\ \bibinfo {author} {\bibfnamefont {J.}~\bibnamefont {Usukura}},\
  }\href@noop {} {\bibfield  {journal} {\bibinfo  {journal} {Few-Body Systems}\
  }\textbf {\bibinfo {volume} {24}},\ \bibinfo {pages} {81} (\bibinfo {year}
  {1998})}\BibitemShut {NoStop}%
\bibitem [{\citenamefont {Muolo}\ \emph
  {et~al.}(2018{\natexlab{a}})\citenamefont {Muolo}, \citenamefont {M\'atyus},\
  and\ \citenamefont {Reiher}}]{Muolo2018b}%
  \BibitemOpen
  \bibfield  {author} {\bibinfo {author} {\bibfnamefont {A.}~\bibnamefont
  {Muolo}}, \bibinfo {author} {\bibfnamefont {E.}~\bibnamefont {M\'atyus}}, \
  and\ \bibinfo {author} {\bibfnamefont {M.}~\bibnamefont {Reiher}},\ }\href
  {\doibase 10.1063/1.5050462} {\bibfield  {journal} {\bibinfo  {journal}
  {J.\,Chem.\,Phys.}\ }\textbf {\bibinfo {volume} {149}},\ \bibinfo {pages}
  {184105} (\bibinfo {year} {2018}{\natexlab{a}})}\BibitemShut {NoStop}%
\bibitem [{\citenamefont {Suzuki}\ and\ \citenamefont
  {Varga}(1998)}]{suzukivarga}%
  \BibitemOpen
  \bibfield  {author} {\bibinfo {author} {\bibfnamefont {Y.}~\bibnamefont
  {Suzuki}}\ and\ \bibinfo {author} {\bibfnamefont {K.}~\bibnamefont {Varga}},\
  }\href@noop {} {\emph {\bibinfo {title} {Stochastic Variational Approach to
  Quantum-Mechanical Few-Body Problems}}}\ (\bibinfo  {publisher}
  {Springer-Verlag, Berlin},\ \bibinfo {year} {1998})\BibitemShut {NoStop}%
\bibitem [{\citenamefont {Simmen}\ \emph {et~al.}(2013)\citenamefont {Simmen},
  \citenamefont {M\'atyus},\ and\ \citenamefont {Reiher}}]{Benjamin2013}%
  \BibitemOpen
  \bibfield  {author} {\bibinfo {author} {\bibfnamefont {B.}~\bibnamefont
  {Simmen}}, \bibinfo {author} {\bibfnamefont {E.}~\bibnamefont {M\'atyus}}, \
  and\ \bibinfo {author} {\bibfnamefont {M.}~\bibnamefont {Reiher}},\
  }\href@noop {} {\bibfield  {journal} {\bibinfo  {journal} {Mol. Phys.}\
  }\textbf {\bibinfo {volume} {111}},\ \bibinfo {pages} {2086} (\bibinfo {year}
  {2013})}\BibitemShut {NoStop}%
\bibitem [{\citenamefont {Muolo}\ \emph
  {et~al.}(2018{\natexlab{b}})\citenamefont {Muolo}, \citenamefont {M\'atyus},\
  and\ \citenamefont {Reiher}}]{Muolo2018a}%
  \BibitemOpen
  \bibfield  {author} {\bibinfo {author} {\bibfnamefont {A.}~\bibnamefont
  {Muolo}}, \bibinfo {author} {\bibfnamefont {E.}~\bibnamefont {M\'atyus}}, \
  and\ \bibinfo {author} {\bibfnamefont {M.}~\bibnamefont {Reiher}},\ }\href
  {\doibase 10.1063/1.5009465} {\bibfield  {journal} {\bibinfo  {journal}
  {J.\,Chem.\,Phys.}\ }\textbf {\bibinfo {volume} {148}},\ \bibinfo {pages}
  {084112} (\bibinfo {year} {2018}{\natexlab{b}})}\BibitemShut {NoStop}%
\bibitem [{\citenamefont {Kinghorn}(1996)}]{Kinghorn1996}%
  \BibitemOpen
  \bibfield  {author} {\bibinfo {author} {\bibfnamefont {D.~B.}\ \bibnamefont
  {Kinghorn}},\ }\href {\doibase
  10.1002/(SICI)1097-461X(1996)57:2<141::AID-QUA1>3.0.CO;2-Y} {\bibfield
  {journal} {\bibinfo  {journal} {Int. J. Quantum Chem.}\ }\textbf {\bibinfo
  {volume} {57}},\ \bibinfo {pages} {141} (\bibinfo {year} {1996})}\BibitemShut
  {NoStop}%
\bibitem [{\citenamefont {Frost}(1967)}]{Frost1967}%
  \BibitemOpen
  \bibfield  {author} {\bibinfo {author} {\bibfnamefont {A.~A.}\ \bibnamefont
  {Frost}},\ }\href {\doibase 10.1063/1.1701524} {\bibfield  {journal}
  {\bibinfo  {journal} {J.\,Chem.\,Phys.}\ }\textbf {\bibinfo {volume} {47}},\
  \bibinfo {pages} {3707} (\bibinfo {year} {1967})}\BibitemShut {NoStop}%
\bibitem [{\citenamefont {Rose}(1957)}]{Rose:AngularMomentum}%
  \BibitemOpen
  \bibfield  {author} {\bibinfo {author} {\bibfnamefont {M.~E.}\ \bibnamefont
  {Rose}},\ }\href@noop {} {\emph {\bibinfo {title} {Elementary Theory of
  Angular Momentum}}}\ (\bibinfo  {publisher} {Wiley, New York},\ \bibinfo
  {year} {1957})\BibitemShut {NoStop}%
\bibitem [{\citenamefont {Cafiero}\ and\ \citenamefont
  {Adamowicz}(2002)}]{Adamowicz2002_FECG}%
  \BibitemOpen
  \bibfield  {author} {\bibinfo {author} {\bibfnamefont {M.}~\bibnamefont
  {Cafiero}}\ and\ \bibinfo {author} {\bibfnamefont {L.}~\bibnamefont
  {Adamowicz}},\ }\href {\doibase 10.1063/1.1457435} {\bibfield  {journal}
  {\bibinfo  {journal} {J.\,Chem.\,Phys.}\ }\textbf {\bibinfo {volume} {116}},\
  \bibinfo {pages} {5557} (\bibinfo {year} {2002})}\BibitemShut {NoStop}%
\bibitem [{\citenamefont {Chaudhry}\ and\ \citenamefont
  {Zubair}(1994)}]{GenIncompGamma_1994}%
  \BibitemOpen
  \bibfield  {author} {\bibinfo {author} {\bibfnamefont {M.}~\bibnamefont
  {Chaudhry}}\ and\ \bibinfo {author} {\bibfnamefont {S.}~\bibnamefont
  {Zubair}},\ }\href {\doibase https://doi.org/10.1016/0377-0427(94)90187-2}
  {\bibfield  {journal} {\bibinfo  {journal} {J. Comp. Appl. Math.}\ }\textbf
  {\bibinfo {volume} {55}},\ \bibinfo {pages} {99 } (\bibinfo {year}
  {1994})}\BibitemShut {NoStop}%
\bibitem [{\citenamefont {Chaudhry}\ \emph {et~al.}(1996)\citenamefont
  {Chaudhry}, \citenamefont {Temme},\ and\ \citenamefont
  {Veling}}]{GenIncompGamma_1996}%
  \BibitemOpen
  \bibfield  {author} {\bibinfo {author} {\bibfnamefont {M.}~\bibnamefont
  {Chaudhry}}, \bibinfo {author} {\bibfnamefont {N.}~\bibnamefont {Temme}}, \
  and\ \bibinfo {author} {\bibfnamefont {E.}~\bibnamefont {Veling}},\ }\href
  {\doibase https://doi.org/10.1016/0377-0427(95)00018-6} {\bibfield  {journal}
  {\bibinfo  {journal} {J. Comp. Appl. Math.}\ }\textbf {\bibinfo {volume}
  {67}},\ \bibinfo {pages} {371 } (\bibinfo {year} {1996})}\BibitemShut
  {NoStop}%
\bibitem [{\citenamefont {Rowan}(1990)}]{Subplex}%
  \BibitemOpen
  \bibfield  {author} {\bibinfo {author} {\bibfnamefont {T.}~\bibnamefont
  {Rowan}},\ }\href@noop {} {\emph {\bibinfo {title} {Functional Stability
  Analysis of Numerical Algorithms}}}\ (\bibinfo  {publisher} {Ph.D. thesis,
  Department of Computer Sciences, University of Texas at Austin},\ \bibinfo
  {year} {1990})\BibitemShut {NoStop}%
\bibitem [{\citenamefont {Brent}(2002)}]{Praxis}%
  \BibitemOpen
  \bibfield  {author} {\bibinfo {author} {\bibfnamefont {R.}~\bibnamefont
  {Brent}},\ }\href@noop {} {\emph {\bibinfo {title} {Algorithms for
  Minimization without Derivatives}}}\ (\bibinfo  {publisher} {Dover},\
  \bibinfo {year} {2002})\BibitemShut {NoStop}%
\bibitem [{\citenamefont {Johnson}()}]{nlopt}%
  \BibitemOpen
  \bibfield  {author} {\bibinfo {author} {\bibfnamefont {S.~G.}\ \bibnamefont
  {Johnson}},\ }\href@noop {} {\bibinfo  {journal}
  {http://github.com/stevengj/nlopt}\ }\BibitemShut {NoStop}%
\bibitem [{\citenamefont {Korobov}(2006)}]{H2+_1}%
  \BibitemOpen
\bibfield  {journal} {  }\bibfield  {author} {\bibinfo {author} {\bibfnamefont
  {V.}~\bibnamefont {Korobov}},\ }\href {\doibase 10.1103/PhysRevA.74.052506}
  {\bibfield  {journal} {\bibinfo  {journal} {Phys. Rev. A}\ }\textbf {\bibinfo
  {volume} {74}},\ \bibinfo {pages} {052506} (\bibinfo {year}
  {2006})}\BibitemShut {NoStop}%
\bibitem [{\citenamefont {Pachucki}\ and\ \citenamefont
  {Komasa}(2009)}]{Pachucki2009}%
  \BibitemOpen
  \bibfield  {author} {\bibinfo {author} {\bibfnamefont {K.}~\bibnamefont
  {Pachucki}}\ and\ \bibinfo {author} {\bibfnamefont {J.}~\bibnamefont
  {Komasa}},\ }\href@noop {} {\bibfield  {journal} {\bibinfo  {journal}
  {J.\,Chem.\,Phys.}\ }\textbf {\bibinfo {volume} {130}},\ \bibinfo {pages}
  {164113} (\bibinfo {year} {2009})}\BibitemShut {NoStop}%
\end{thebibliography}
\newcommand{\Aa}[0]{Aa}

\end{document}